# Title: Van Hove Singularity-Driven Topological Magnetism in Twisted $MoTe_2$

**Authors:** Heonjoon Park[1], Julian Stewart[2], Xiao-Wei Zhang[2], Taige Wang[3,4], Canxun Zhang[5], Evgeny Redekop[5], Jiaqi Cai[1], Weijie Li[1], Eric Anderson[1], Takashi Taniguchi[6], Kenji Watanabe[7], Jiun-Haw Chu[1], David Cobden[1], Andrea Young[5], Liang Fu[4], Ting Cao[2], Di Xiao[1,2], Xiaodong Xu[1,2*]

**Affiliations:**

[1]Department of Physics, University of Washington, Seattle, Washington 98195, USA

[2]Department of Materials Science and Engineering, University of Washington, Seattle, Washington 98195, USA

[3]Department of Physics, Harvard University, Cambridge, Massachusetts 02138, USA

[4]Department of Physics, Massachusetts Institute of Technology, Cambridge, Massachusetts 02139, USA

[5]Department of Physics, University of California - Santa Barbara, Santa Barbara, California 93106, USA

[6]Research Center for Materials Nanoarchitectonics, National Institute for Materials Science, 1-1 Namiki, Tsukuba 305-0044, Japan

[7]Research Center for Electronic and Optical Materials, National Institute for Materials Science, 1-1 Namiki, Tsukuba 305-0044, Japan

*Corresponding author. Email: xuxd@uw.edu

**Abstract: Van Hove singularities (vHSs) strongly amplify electron interactions and can stabilize correlated phases in topological bands. Here we report signatures of topological magnetism in large-angle twisted bilayer $MoTe_2$ driven by the interplay of vHSs, strong correlations, and valley topology. In a 4.8° device, electrostatic tuning to a vHS produces a spontaneous anomalous Hall hot spot near $\nu \approx -1$. Combined transport and reflective magnetic circular dichroism measurements indicate that this regime is not governed by magnetization alone, but instead emerges from a correlated intervalley-coherent antiferromagnetic state that evolves with doping into a canted phase. With increasing magnetic field, the Hall response develops an additional finite-field component consistent with a topological Hall effect from a noncoplanar spin texture, before transitioning into a $C = -1$ Chern insulator. Our results establish tunable vHSs in moiré topological bands as a route to chiral magnetism and engineering topological phase transitions.**

**Main Text:**

Saddle points in Bloch bands give rise to van Hove singularities (vHSs) (*1*), where the density of states diverges and interaction effects are strongly enhanced (*2-4*). In conventional bulk materials, these singularities are typically far from the Fermi level and play a limited role in accessible phases. In two-dimensional moiré systems, however, twist-angle engineering can place vHSs within reach of the Fermi level, allowing them to be tuned in situ by electrostatic gating.. Recent experiments in twisted transition metal dichalcogenides, particularly twisted $WSe_2$, have demonstrated that vHSs can drive correlated phases, including superconductivity and magnetism (*5-12*). These results highlight the importance of vHS-enhanced interactions, but the role of vHSs in generating topological states remains largely unexplored. In particular, the momentum-space structure near a vHS can favor unconventional magnetic instabilities and interaction-driven Fermi surface reconstruction (*13, 14*), suggesting a distinct route to topological phases beyond the flat-band paradigm.

Twisted $MoTe_2$ provides a natural platform to explore this regime. Its moiré bands carry valley-contrasting Chern numbers and can be tuned by an out-of-plane electric field, which controls both bandwidth and lattice geometry (*15*). At small twist angles, the system realizes integer and fractional quantum anomalous Hall states characteristic of the flat-band limit (*16-23*). At larger twist angles, the moiré bands become more dispersive, driving the system away from the flat-band limit and into an intermediate-coupling regime. In this regime, van Hove singularities emerge within the valence bands and can be tuned across the Fermi level, providing direct access to vHS-driven instabilities within a topological band structure. Theory further predicts that such vHSs can promote spin- and valley-polarized instabilities (*14, 24-26*) and a range of correlated phases, including magnetic order (*27-31*), chiral/topological superconductivity (*32-40*), and quantum spin liquid states (*41, 42*).

Here we show that vHSs in twisted $MoTe_2$ drive a correlated magnetic regime with topological transport signatures. Combining transport and magnetic circular dichroism measurements, we find evidence consistent with an intervalley-coherent antiferromagnetic ground state near $\nu \approx -1$, which evolves with doping into a canted phase exhibiting a spontaneous anomalous Hall effect, and further transitions under magnetic field into a $C = -1$ Chern insulator. The Hall response additionally contains a finite-field component that does not track the magnetization, consistent with a topological Hall effect arising from a noncoplanar spin texture with finite scalar spin chirality. Supported by Hartree–Fock calculations, these results establish a unified picture in which vHS-driven Fermi surface reconstruction promotes chiral magnetism and generates an emergent gauge field that governs electronic transport.

We first consider the evolution of the electronic structure in twisted $MoTe_2$ as a function of twist angle. As shown in Fig. 1A, the moiré superlattice consists of three high symmetry stacking configurations, denoted as MX, XM, and MM (M = Mo, X = Te). The resulting valence bands (Fig. 1B and Supplementary Fig. S1) carry opposite Chern numbers (±1) in the K and K′ valleys. An out-of-plane electric field ($D/\varepsilon_0$) lifts the layer degeneracy by polarizing the layer pseudospin and reshaping the band structure. In particular, the first valence band hosts saddle points that give rise to vHSs (Fig. 1B), whose position in filling factor $\nu$ shift with electric field, as reflected in the DOS (Fig. 1C and Supplementary Fig. S2). We now turn to transport to track how these features evolve with twist angle (see Supplementary Fig. S3). Unless otherwise specified, all transport measurements are performed at temperatures below 100 mK.

Figures 1D-F displays the symmetrized longitudinal ($R_{xx}$) and antisymmetrized Hall ($R_{xy}$) resistances, as functions of $\nu$ and $D/\varepsilon_0$ at a magnetic field of $\mu_0 H = \pm 100$ mT for devices with different twist angles. In the 4.2° device (Fig. 1D), a well-developed QAH phase emerges near $\nu = -1$ at low displacement fields ($|D/\varepsilon_0| < 150$ mV/nm), characterized by a vanishing longitudinal resistance and a Hall resistance quantized to $h/e^2$. The magnetic field-dependent measurements at $\nu = -1$, $D/\varepsilon_0 = 100$ mV/nm (Fig. 1G) further confirm this behavior, showing a quantized Hall plateau accompanied by suppressed $R_{xx}$ at zero magnetic field. As the displacement field increases beyond $|D/\varepsilon_0| \approx 150$ mV/nm, the anomalous Hall response weakens and eventually disappears, giving way to a strongly insulating phase. This regime is likely associated with a topologically trivial Mott-like insulating state (see Supplementary Fig. S4 for detailed electric field dependence). Overall, the phenomenology of the integer QAH phase in the 4.2° device closely resembles that observed in smaller-angle systems (*18, 19*).

As the twist angle becomes larger, the Chern bands become more dispersive (Supplementary Fig. S2), weakening interaction-driven gaps and destabilizing the QAH phase. In the 4.5° device (Fig. 1E), both the QAH state and spontaneous ferromagnetism are absent near $\nu = -1$ and $D/\varepsilon_0 \approx 0$, where a non-ferromagnetic metallic state is seen instead. However, a distinct behavior appears at intermediate displacement fields, 50 mV/nm < $D/\varepsilon_0$ < 250 mV/nm (Supplementary Fig. S4, S5). In this regime, $R_{xx}$ exhibits a pronounced minimum, while $R_{xy}$ develops a large anomalous Hall signal of approximately 8 kΩ with clear hysteresis (Fig. 1H). With increasing magnetic field, $R_{xy}$ evolves smoothly toward quantization while $R_{xx}$ vanishes above ~9 T, consistent with a Chern number $C = -1$ from the Středa relation (Supplementary Fig. S5). These observations identify the zero-field phase as an incipient Chern insulator (Fig. 1H, inset), in which the gap is fully developed only with the assistance of a Zeeman coupling. The emergence of this state reflects the interplay of interaction, band topology, and the enhanced density of states near the van Hove singularity. This response is similar to vHS-driven instabilities reported in twisted $WSe_2$ (*5, 6, 8, 9*), but unlike $WSe_2$, $tMoTe_2$ hosts an isolated lowest moiré miniband with a comparatively flatter bandwidth and reduced band mixing, resulting in clear Chern insulating signatures (*43*).

At a larger twist angle of 4.8°, the incipient Chern insulating phase is replaced by multiple resistive states that appear near $\nu = -1$ at intermediate displacement fields around ±200 mV/nm, forming AHE "hot spots" in Fig. 1F (Supplementary Fig. S4, S6). These states exhibit hysteresis upon magnetic field cycling (Fig. 1I), but the AHE signal (~1 kΩ) is much smaller than $R_{xx}$ (~10 kΩ). A similar resistive feature without AHE signal is observed near $\nu = -1/2$ at finite displacement field. Correlated insulating states also develop near $D/\varepsilon_0 \approx 0$ at $\nu = -1/3$, $-1/4$, $-1/6$, and $-1/7$, which are non-magnetic (Supplementary Fig. S7).

Given this behavior, we focus on the 4.8° device hereafter. The longitudinal resistance near $\nu = -1$ exhibits a complex structure as a function of $\nu$ and $D/\varepsilon_0$ (Fig. 2A). Three higher-resistance branches emanate from a central resistive pocket, forming a characteristic fishtail-like pattern. Similar bifurcating features have been reported in 5° $tWSe_2$, where two branches disperse toward higher $\nu$ and $D/\varepsilon_0$ (*6, 8, 9*). In contrast, an additional lower branch is observed here, shifting toward lower $D/\varepsilon_0$ with increasing $\nu$, which appears unique to $tMoTe_2$. This branch can also be seen at smaller twist angles (Figs. 1D, E).

Along these branches, $R_{xx}$ develops pronounced maxima, indicating enhanced scattering. This behavior is consistent with a reorganization of the electronic structure near van Hove singularities, where interaction effects can reconstruct the Fermi surface, split the singularities, and open partial gaps (Fig. 3F) (*26*). In contrast, the Hall response occupies a much narrower region in phase space and displays a comparatively simple structure. As shown in Fig. 2B, the anomalous Hall signal is centered near $\nu \approx -0.94$ and spans only a limited range of electric field. The distinct behaviors of $R_{xx}$ and $R_{xy}$ suggest the presence of multiple competing states near the vHS. To directly probe their magnetic origin, we employ magnetic circular dichroism measurements.

Reflective magnetic circular dichroism (RMCD) measures valley polarization via the differential reflectance between left- and right-circularly polarized light (see Methods). Figures 2D-F show the RMCD signal as a function of $\nu$ and $D/\varepsilon_0$ at selected magnetic fields $\mu_0 H$. At zero field, the RMCD map closely tracks the anomalous Hall signal (Fig. 1F), with a hotspot in the same region of phase space. With increasing magnetic field, however, the RMCD response extends beyond this hotspot and progressively follows the vHS branches. Above ~50 mT (Supplementary Fig. S8), the signal becomes broadly enhanced along the entire vHS arc. This behavior is consistent with an enhanced magnetic susceptibility arising from the large DOS near the vHS (Figs. 1C and 2F), which promotes spin polarization under weak fields. In contrast, the anomalous Hall signal remains confined to the hotspot region and shows little field dependence in this regime (Fig. 1F and Supplementary Figs. S6, S11, S12). The distinct phase boundaries of the RMCD and $R_{xy}$ are inconsistent with a simple Stoner mechanism as the origin of the hysteretic AHE, since Stoner magnetism is governed by a single order parameter, the magnetization.

To further characterize the magnetic interactions near $\nu = -1$, we measure RMCD while cycling the magnetic field. Clear hysteresis is seen for $-0.95 < \nu < -0.88$ (Fig. 2D inset, Supplementary Fig. S9), and is used to extract the Curie temperature $T_C$ as a function of filling (Fig. 2G). We also use the low-field slope of the RMCD signal as an approximation of the magnetic susceptibility $\chi_{MCD}$, following an established procedure (*15, 44, 45*). The temperature dependence of $\chi_{MCD}$ is shown in Fig. 2G. By fitting $1/\chi$ versus $T$ to the Curie–Weiss form (Supplementary Fig. S10), we extract the Curie–Weiss temperature $T_{CW}$ vs. filling factor (Fig. 2H). Near $\nu = -1$, $T_{CW}$ is negative, indicating antiferromagnetic interactions between local moments. This regime is consistent with a predicted intervalley-coherent antiferromagnetic (IVC-AFM) state (Fig. 3F) which is characterized by a 120° in-plane spin pattern in real space (*26, 27, 29, 46, 47*). Upon doping away from $\nu = -1$, $T_{CW}$ becomes positive and coincides with the onset of RMCD hysteresis, suggesting the emergence of a noncoplanar canted AFM state (Fig. 3G) (*26*). Taken together, these results motivate the tentative phase diagram shown in Fig. 2C, delineating vHS features and two magnetic regimes: IVC-AFM near $\nu = -1$ and a doped canted AFM phase.

We further support this phase diagram by examining transport as a function of magnetic field and filling near $\nu = -1$. Figures 3A-B show $R_{xx}$ and $R_{xy}$, respectively, while Fig. 3C isolates the hysteretic component of $R_{xy}$ from field-up and field-down sweeps, revealing multiple competing phases in close proximity. At $\nu \approx -0.93$, a pronounced AHE is observed, indicating spontaneous time-reversal symmetry breaking. As $\nu$ approaches $-1$, the AHE is suppressed, giving way to a state with negligible $R_{xy}$ at low fields. Upon increasing the magnetic field beyond a $\nu$-dependent threshold (up to ~1 T), the system undergoes a transition into a high-$R_{xy}$ state. This transition is marked by a triangular region of enhanced $R_{xx}$ (Fig. 3A) and a sharp step in $R_{xy}$ (Fig. 3B), clearly delineating a phase boundary.

The nature of this field-induced state becomes clear at higher fields. Figure 3D shows $R_{xx}$ up to 11 T, where a pronounced resistance minimum develops near $\nu = -1$ once the transition is crossed. The trajectory of this minimum follows a Streda slope of $-1$, identifying a Chern insulating phase with $C = -1$ (Supplementary Figs. S11-13). Consistently, Fig. 3E shows $|R_{xy}|$ increasing with magnetic field and approaching $h/e^2$ near $\nu = -1$. Together, these results demonstrate that the applied magnetic field drives the system from a low-field, valley-imbalanced state into a fully valley-polarized Chern-insulating state. Through Zeeman coupling, spins align out of plane, lifting valley degeneracy in the Chern band and stabilizing a $C = -1$ Chern insulator, as summarized schematically in Fig. 3H.

In order to examine the AFM behavior, we plot line traces of $R_{xx}$ and $R_{xy}$ as a function of $\mu_0 H$ over a small field range at selected $\nu$ in Figs. 4A-B along fixed $D/\varepsilon_0$ = 194 mV/nm (see also Supplementary Fig. S14). Two distinct transport responses emerge. In the first, $R_{xy}$ is zero at $\mu_0 H$ = 0, increases gradually with field, and then rises sharply beyond a $\nu$-dependent threshold. This behavior is consistent with an antiferromagnetic ground state (*9*): at low fields, spins cant continuously, producing a weak Hall response, while above a critical field they fully align, coinciding with the onset of the field-induced Chern insulator discussed in Fig. 3. The corresponding RMCD traces (Fig. 4C) follow the same evolution, vanishing at zero field, increasing smoothly with $\mu_0 H$ due to canting, and saturating at higher fields when the spins are fully polarized. This regime corresponds to the triangular region of enhanced $R_{xx}$ identified in Fig. 3A near $\nu \approx -1$. The temperature dependence of the low-field Hall response (Supplementary Fig. S15) shows an onset scale around ~3 K, consistent with antiferromagnetic ordering.

A qualitatively different behavior appears in the region with finite AHE. Here, $R_{xy}$ exhibits a pronounced peak at finite magnetic field, most clearly near $\nu \approx -0.93$ (Figs. 4B), resembling a topological Hall-like response (*48-51*). A concomitant peak in $R_{xx}$, followed by negative magnetoresistance at higher field, is consistent with the suppression of spin-disorder or texture-related scattering as the spins become progressively aligned. This feature cannot be attributed to straightforward magnetic domain dynamics (*52-55*). The coercive field associated with magnetization reversal is below ~10 mT (Fig. 3C and inset of Fig. 4B), far smaller than the ~100 mT scale at which the peak in $R_{xy}$ develops (black arrows in Fig. 4B). Moreover, the absence of a corresponding feature in the RMCD signal (Fig. 4C), which tracks the net magnetization, further rules out a simple ferromagnetic origin. Further scanning nanoSQUID-on-tip measurements reveal a largely homogeneous mesoscopic magnetization across the device channels within the canted IVC-AFM regime (Supplementary Fig. S16), further excluding mesoscopic domain inhomogeneity as the source.

Combined with the non-Stoner character of the AHE established in Fig. 2, these results indicate that the Hall response near $\nu \approx -1$ is governed by the field-driven evolution of a correlated IVC state rather than by net magnetization alone. Self-consistent Hartree–Fock (HF) calculations indicate that, at exactly $\nu = -1$ and large displacement field, the fully valley-polarized limit is an interaction-reconstructed $C = 0$ insulator even though the corresponding single-particle band carries $C = -1$, implying that the Hall hotspot must arise before full polarization is reached. At zero magnetic field, the ground state is predominantly intervalley coherent and nearly time-reversal symmetric, so the Berry-curvature contributions from opposite valleys largely cancel and $R_{xy}$ remains small. A small residual valley polarization naturally accounts for the weak remanent RMCD signal. For fillings slightly above $\nu = -1$, increasing magnetic field acts primarily through Zeeman coupling, which continuously cants the order parameter away from the IVC limit toward a valley polarized state. As shown in Fig. 4D-E, the intervalley-coherent order parameter $\underline{U}_v$ (1) (red) decreases while the valley polarization $P_v$ (blue) increases, whereas $\sigma_{xy}$ first rises sharply as the cancellation between opposite-valley Berry curvatures is lifted, and then decreases again on approaching the fully polarized regime. This identifies the Hall hotspot with an intermediate-field canted state. Details of the Hartree-Fock calculations are provided in the Supplementary Materials. Notably, the Hall conductivity $\sigma_{xy}$ extracted from the measured $R_{xx}$ and $R_{xy}$ shows a qualitatively similar field dependence (Fig. 4F), in qualitative agreement with the Hartree-Fock result.

A natural real-space picture of this intermediate-field regime is a non-coplanar canted IVC-AFM texture with finite scalar spin chirality. As illustrated in Fig. 4G, the scalar spin chirality $S_i \cdot (S_j \times S_k)$ breaks time-reversal symmetry and, through the solid angle subtended by the spin texture, imparts a Berry phase to itinerant electrons (Fig. 4I inset). Chiral spin textures of this type have been extensively studied in frustrated and strongly correlated systems, where they produce unconventional transport responses including anomalous Hall and topological Hall effects (*56-60*). In the present case, the finite out-of-plane spin component gives rise to the remanent RMCD signal, while the non-coplanar texture generates an emergent magnetic field that deflects carriers according to spin texture rather than net magnetization alone. This mechanism is closely related to geometric spin-orbit coupling induced by spin textures (*61*), and provides a natural interpretation of the enhanced Hall response as a van-Hove-singularity-induced topological Hall effect.

Experimentally, the Hall response can be decomposed phenomenologically as $R_{xy} = R_{LH}(B) + R_{AH}(M) + R_{TH}$, where $R_{LH}$ is the normal Hall contribution proportional to $B$, $R_{AH}$ denotes the part that follows the RMCD background, with RMCD used here as a proxy for $M$, and $R_{TH}$ is the residual contribution. Figure 4H shows this decomposition at $\nu = -0.93$. The measured signal (blue) is separated into a background estimated from RMCD and a residual component (black) identified as $R_{TH}$. The peak in $R_{TH}$ occurs at finite magnetic field and does not follow the magnetization, consistent with a topological origin. The extracted $R_{TH}$ versus filling (Fig. 4I) shows a pronounced maximum near $\nu \approx -0.94$, coincident with the AHE hotspot. Together, these results support a picture in which the vHS promotes competing valley orders, while Zeeman coupling stabilizes a chiral intermediate state with maximal Hall response, accounting for both the AHE hotspot and the additional topological Hall contribution. At still higher field and exactly $\nu = -1$, a competing $C$= -1 Chern insulator remains close in energy and is eventually stabilized by magnetic field, driving a first-order transition into the Chern insulating state.

While small-angle twisted $MoTe_2$ realizes integer and fractional quantum anomalous Hall states as flat-band analogues of Landau levels, our results show that increasing twist angle drives the system into a qualitatively different regime where finite bandwidth and non-uniform quantum geometry become essential. Combining transport and RMCD measurements, we identify a correlated phase near $\nu \approx -1$ tied to van Hove singularities, where the enhanced density of states promotes spin-valley instabilities and unconventional magnetic order. The data support an intervalley-coherent antiferromagnetic ground state that evolves with doping into a canted phase with a spontaneous anomalous Hall response, and with magnetic field into a $C = -1$ Chern insulator. In addition, the Hall response contains a distinct contribution beyond conventional anomalous Hall physics, consistent with a topological Hall effect arising from a chiral spin texture. Supported by Hartree–Fock calculations, these observations point to a noncoplanar IVC-AFM state with finite scalar spin chirality near the van Hove singularities, generating an emergent gauge field that directly impacts charge transport. More broadly, our results demonstrate that in moiré Chern bands, the interplay of band dispersion, topology, and interactions can stabilize competing and tunable correlated phases, where chiral spin order emerges from a doped IVC-AFM background and gives rise to topological transport signatures distinct from the flat-band limit.

**Note added:** During the preparation of this manuscript, two related studies (*62, 63*) appeared reporting electrical transport measurements in twisted $MoTe_2$ in a similar 4° to 5° regime.

**Funding:** This work was mainly supported by US Department of Energy (DOE), Basic Energy Science (BES) under award DE-SC0018171. The fabrication and measurement are supported partially by the Vannevar Bush Faculty Fellowship (Award number N000142512047). Bulk $MoTe_2$ crystal growth and characterization is supported by Programmable Quantum Materials, an Energy Frontier Research Center funded by DoE BES under award DE-SC0019443. Sample fabrication was partially performed by using instrumentation supported by the US National Science Foundation through the UW Molecular Engineering Materials Center (MEM·C), a Materials Research Science and Engineering Center (DMR-2308979). Work at UCSB was supported by the Army Research Office (Award number W911NF-20-2-0166). A.Y. acknowledges support from the Gordon and Betty Moore Foundation EPIQS program under award GBMF9471. D.X. acknowledges support from DOE BES under the award DE-SC0012509. E.A. acknowledges support from the National Science Foundation Graduate Research Fellowship Program under Grant No. DGE-2140004. K.W. and T.T. acknowledge support from the JSPS KAKENHI (Grant Numbers 21H05233 and 23H02052), the CREST (JPMJCR24A5), JST and World Premier International Research Center Initiative (WPI), MEXT, Japan. X.X. and J.-H.C. acknowledge support from the State of Washington funded Clean Energy Institute.

**Author contributions:** H.P. and X.X. conceived the project. H.P. fabricated the devices. H.P. performed the transport measurements, with assistance from J.C. J.S., W.L., E.A., performed the magneto-optical measurements. C.Z., E.R., and A.Y. performed the nano-SQUID-on-tip measurements. H.P., J.S., D.C., L.F., T.C., D.X., and X.X. analyzed and interpreted the results. C.H. and J.-H.C. grew and characterized the bulk $MoTe_2$ crystals. X.Z., T.C., and D.X., modeled the band structure. T.W., and L.F. performed the Hartree-Fock calculation. H.P., D.X., and X.X. wrote the paper with input from all authors. All authors discussed the results.

**Competing interests:** Authors declare that they have no competing interests.

**Data and materials availability:** We have included the data in the main text and supplementary materials cited.

**Supplementary Materials**
Materials and Methods
Supplementary Text
Figs. S1 to S16
References (*64-77)*

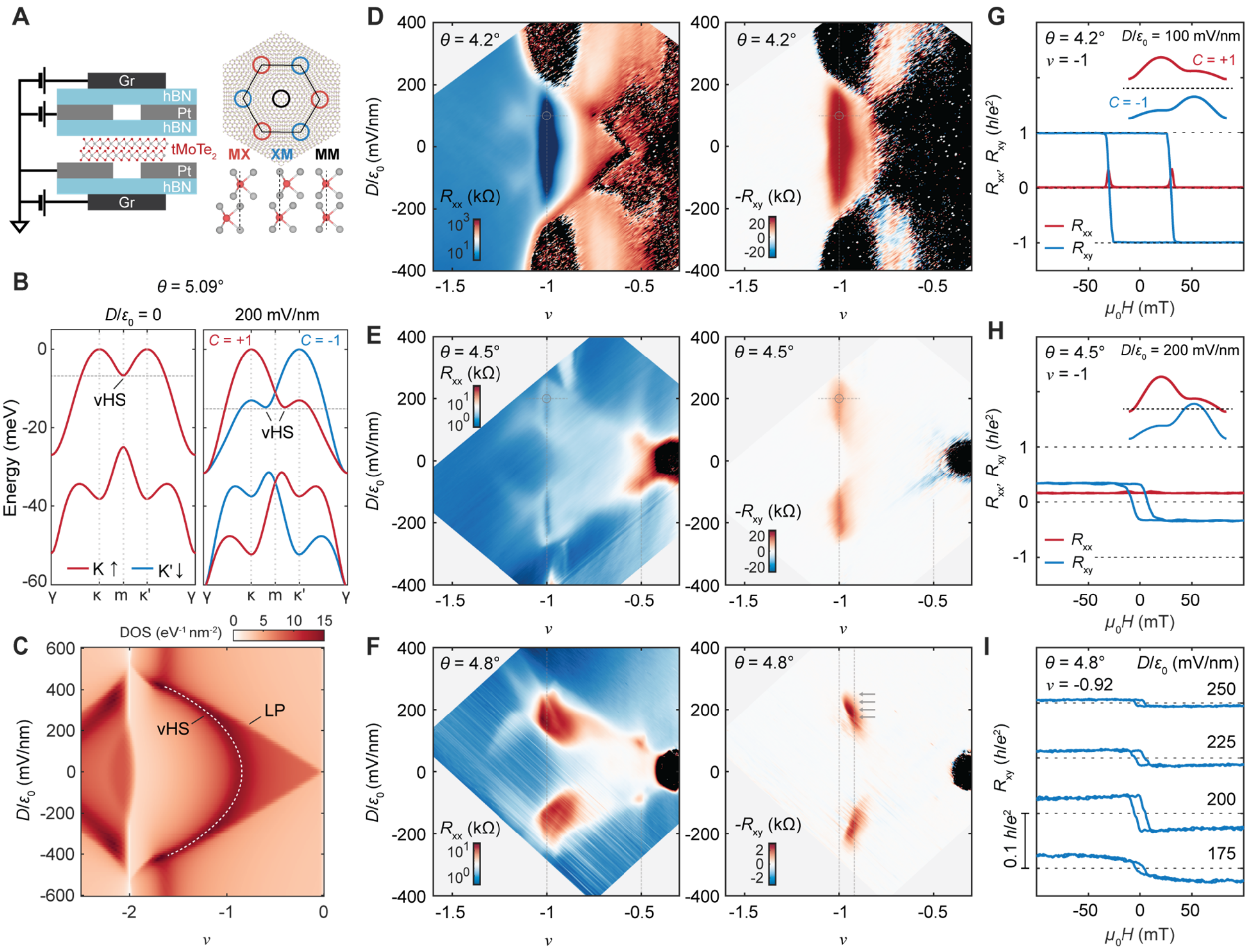

**Fig. 1. Twist angle-tuned topology, magnetism, and van Hove singularities (vHSs) in homobilayer $MoTe_2$.** **A**, Schematic of the dual-gated device (left) and the moiré superlattice of twisted bilayer $MoTe_2$ (right). **B**, Calculated moiré band structure for a 5.09° device at electric fields of $D/\varepsilon_0 = 0$ (left) and 200 mV/nm (right), showing electric-field-tunable vHSs. **C**, Calculated density of states (DOS) as a function of electric field and filling factor $\nu$ at $\theta = 5.09°$. The vHS is marked by the dotted white line, while the layer polarization (LP) boundary is identified by the sharp change in DOS. **D–F,** Longitudinal resistance $R_{xx}$ (left panels) and Hall resistance $R_{xy}$ (right panels) as functions of filling factor $\nu$ and electric field $D/\varepsilon_0$ for devices with twist angles 4.2° (**D**), 4.5° (**E**), and 4.8° (**F**), respectively. The maps show a systematic evolution from a robust Chern insulating regime near $\nu = -1$ at smaller angles to features tied to the vHS at larger angles. **G–I,** Corresponding magnetotransport at fixed points indicated in **D–F**, respectively. **G,** For $\theta = 4.2°$, magnetic field sweeps show a quantized anomalous Hall (QAH) state with clear hysteresis. (Inset: schematic illustration of the moiré Chern bands at finite electric field with spontaneous spin/valley polarization). **H,** At $\theta = 4.5°$, the QAH state near $\nu = -1$ is suppressed, while an incipient Chern insulating state develops near the vHS at finite electric field. (Inset: schematic of partially polarized Chern band at finite $D/\varepsilon_0$) **I,** For $\theta = 4.8°$, $R_{xy}$ versus magnetic field at selected electric fields (offset by 0.1 $h/e^2$ for clarity) highlights anomalous Hall responses concentrated near the vHS and slightly below $\nu = -1$. All measurements are performed below 100 mK. $R_{xx}$ and $R_{xy}$ maps are symmetrized and antisymmetrized with respect to magnetic field at ±100 mT, respectively.

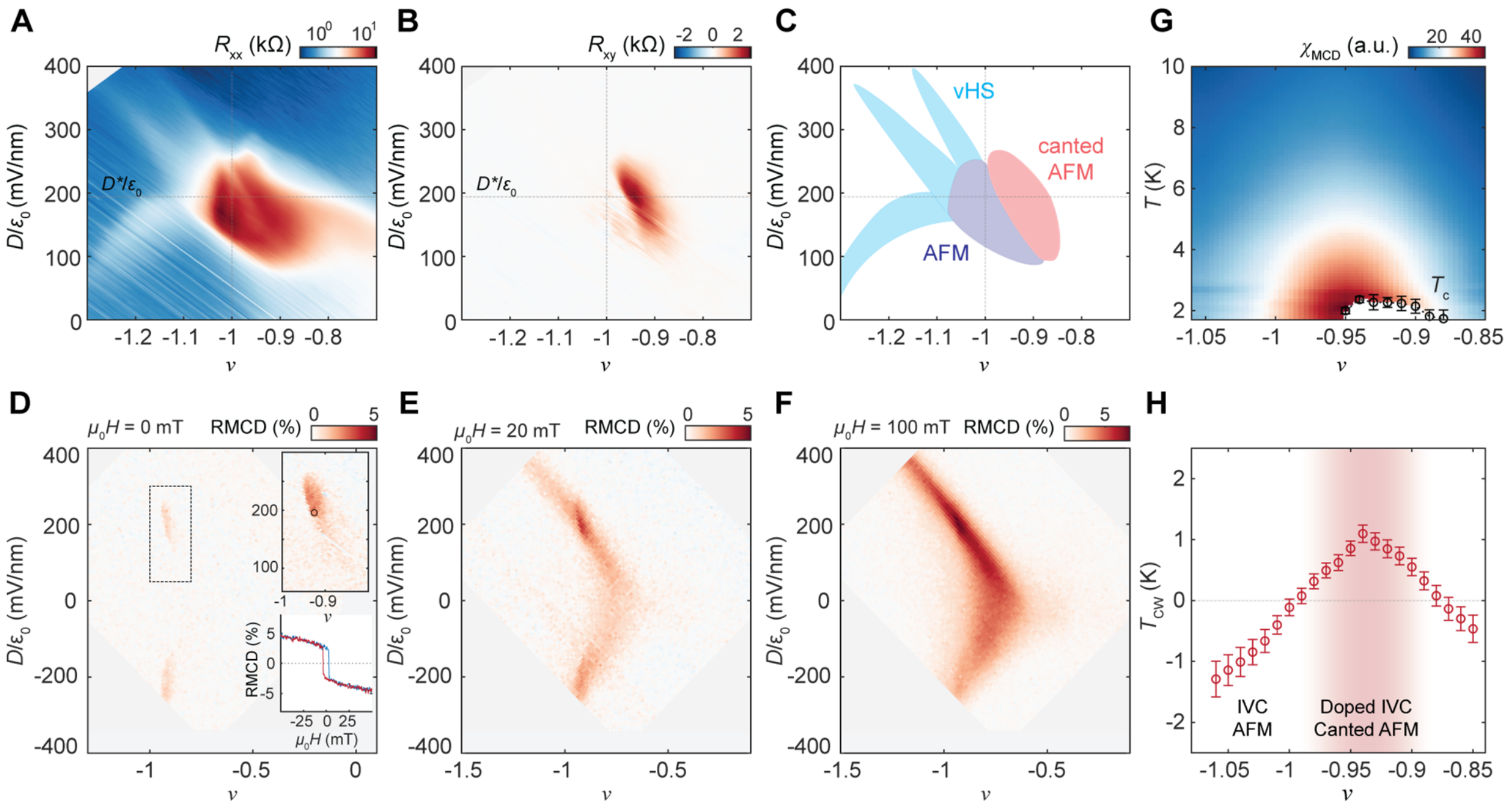

**Figure 2. vHS-driven anomalous Hall effect hot spots in the 4.8° device. A-B**, Zoomed-in map of $R_{xx}$ (**A**) and $R_{xy}$ (**B**) near the vHSs, measured under the same conditions as Fig. 1F. $R_{xx}$ exhibits a complex structure with multiple resistive peaks (see Supplementary Fig. S6 for temperature dependence). The anomalous Hall effect (AHE) emerges at finite electric field and slightly below $\nu = -1$. **C**, Schematic phase diagram near the vHS. **D-F**, Reflective magnetic circular dichroism (RMCD) as a function of $D/\varepsilon_0$ and $\nu$ at magnetic fields $\mu_0 H = 0$ mT (**D**), 20 mT (**E**), and 100 mT (**F**). The RMCD signal indicates that spin polarization rapidly develops with applied magnetic field along the vHS feature. However, the AHE is confined to a localized hotspot (**B**) even at 100 mT, indicating a distinct origin beyond simple Stoner ferromagnetism. The upper inset of (**D**) shows a zoom-in to the RMCD signal near the hotspot while the lower inset in (**D**) shows RMCD as the magnetic field is cycled at the condition marked by the black circle in the upper inset. **G**, Magnetic susceptibility $\chi_{MCD}$ extracted from the low-field RMCD response as a function of temperature and filling factor (see Supplementary Fig. S9, S10). Black circles denote the Curie temperature $T_C$, determined from the disappearance of RMCD hysteresis. Below $T_C$, $\chi_{MCD}$ cannot be reliably extracted. **H**, Extracted Curie–Weiss temperature $T_{CW}$ as a function of $\nu$ (see Methods and Supplementary Fig. S10), indicating antiferromagnetic interactions near $\nu \approx -1$ that evolve into positive $T_{CW}$ upon doping.

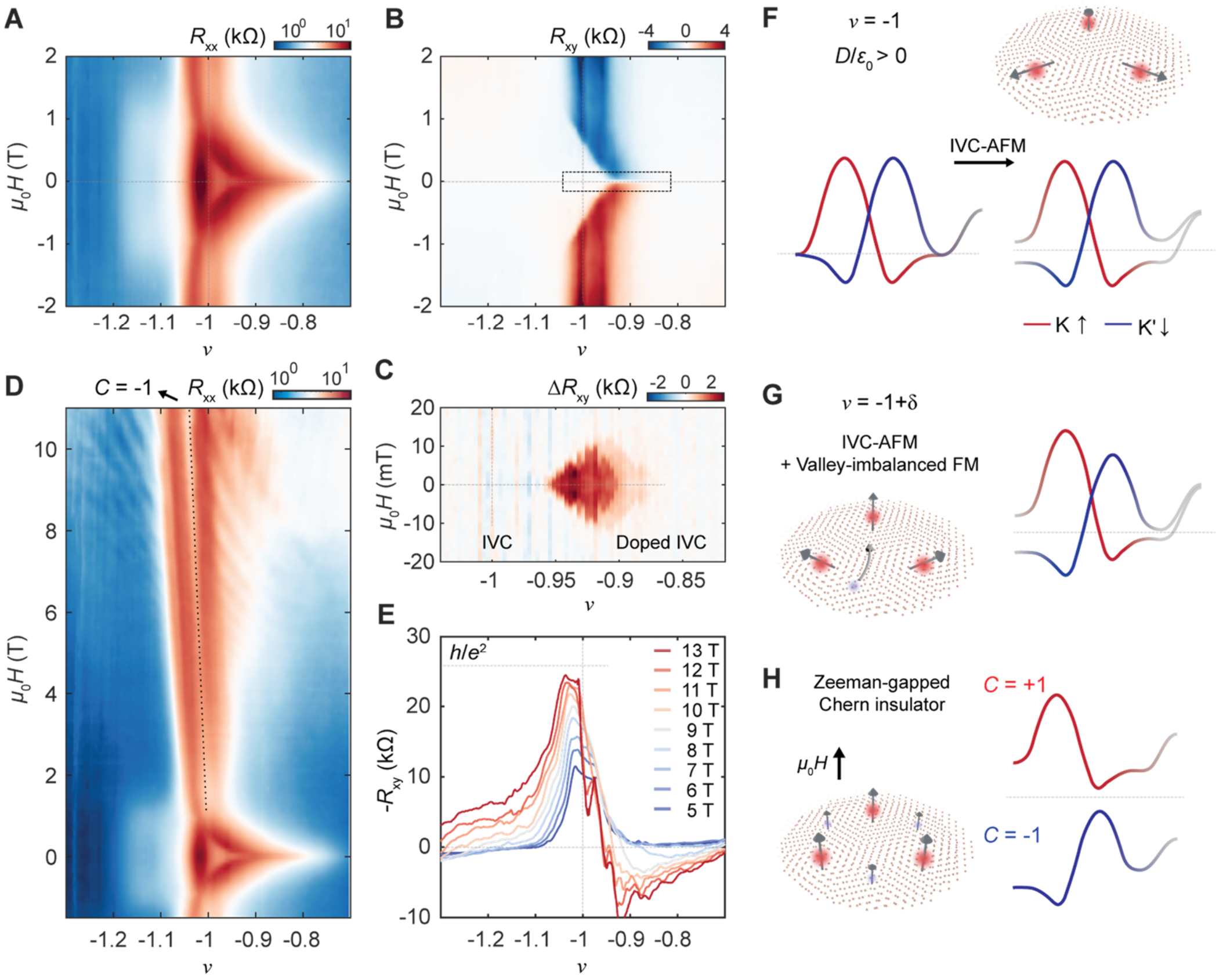

**Figure 3. Magnetic-field-induced Chern insulator emerging from vHS-driven correlated phases.** All data are taken at fixed $D^*/\varepsilon_0$ = 194 mV/nm, shown in Figs. 2A-B. **A-B**, $R_{xx}$ (**A**) and $R_{xy}$ (**B**) as a function of magnetic field and filling factor near $\nu = -1$. **C**, Hysteretic component $\Delta R_{xy}$, defined as the difference between up and down magnetic field sweeps, showing a localized region of enhanced anomalous Hall response in the dotted rectangle in (**B**). **D**, $R_{xx}$ vs $\nu$ extended to high magnetic fields. A pronounced resistance minimum develops, consistent with a magnetic-field-induced Chern insulating state following a Streda slope corresponding to $C = -1$. **E**, $R_{xy}$ as a function of $\nu$ at selected magnetic fields. With increasing field, $R_{xy}$ grows and approaches $h/e^2$. **F-H**, Schematic illustration of correlated phases inferred from transport. At $\nu = -1$ and finite $D/\varepsilon_0$, interaction-driven spin-valley instabilities reconstruct the Fermi surface into an intervalley-coherent antiferromagnetic (IVC-AFM) state with a partial gap near the van Hove singularity, forming a correlated metal (**F**). Upon slight doping (**G**), the reconstructed state develops a valley imbalance, leading to a canted IVC-AFM phase. With increasing out-of-plane magnetic field (**H**), Zeeman coupling gaps the reconstructed bands, resulting in a Chern insulating state with $C = -1$.

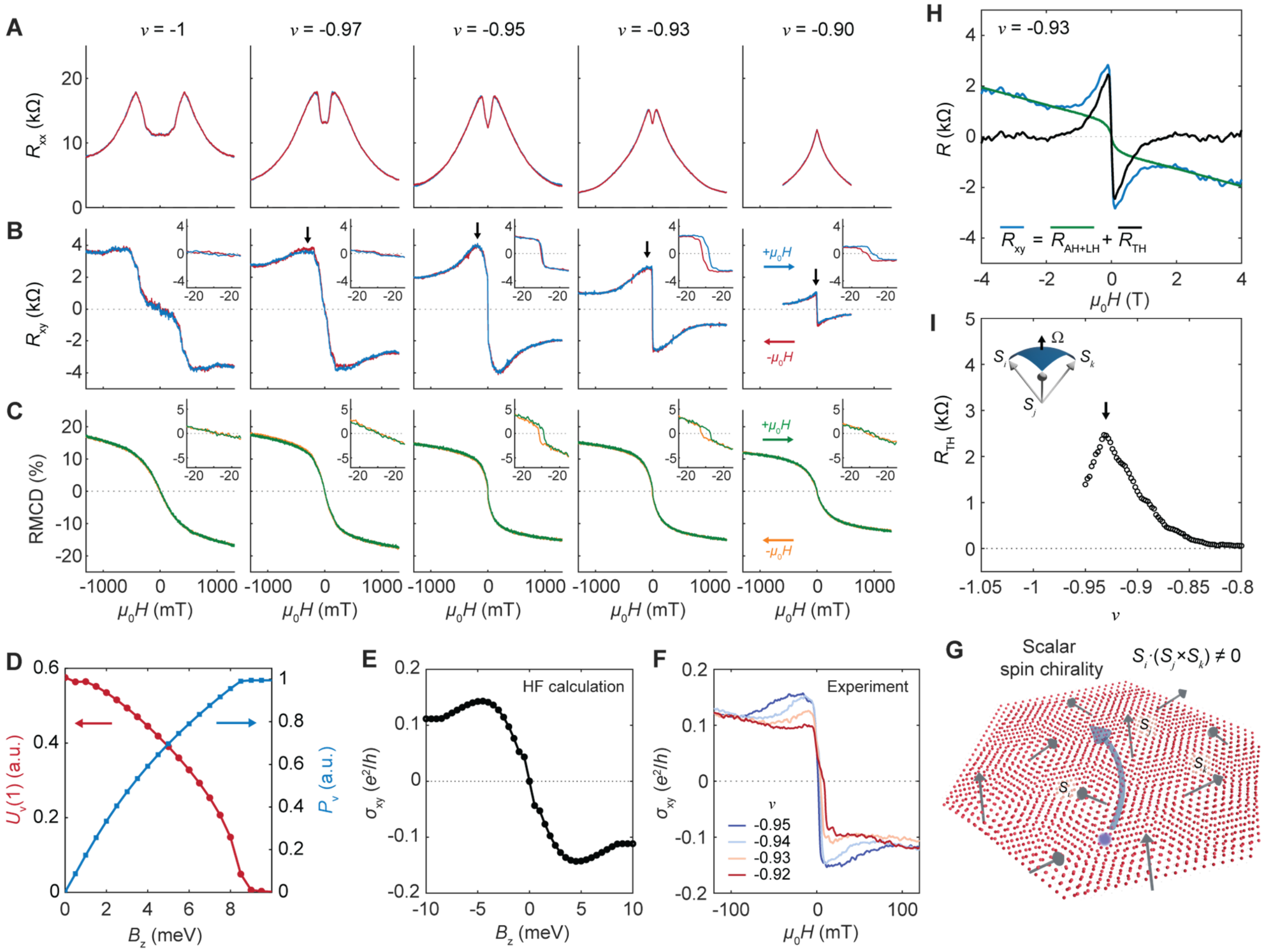

**Figure 4. Topological Hall effect and chiral spin texture.** **A-C**, $R_{xx}$ (**A**), $R_{xy}$ (**B**), and RMCD (**C**) as functions of magnetic field at selected filling factors $\nu$, measured at fixed $D^*/\varepsilon_0$ = 194 mV/nm shown in Figs. 2A-B. Colors indicate the magnetic field sweep direction. Insets show enlarged views of the low-field region, using the same axes and units, to emphasize the hysteresis. Black arrows mark the field at which the topological Hall contribution is largest, around 100 mT, well above the coercive field of the anomalous Hall hysteresis, around 10 mT. **D-E**, Hartree-Fock calculations show the evolution of the intervalley-coherent order parameter $U_v(1)$ and valley polarization $P_v$ with Zeeman field (**D**), together with the corresponding Hall conductivity $\sigma_{xy}$ (**E**). See Supplementary Information for details of the calculation. **F**, Experimentally extracted Hall conductivity $\sigma_{xy}$ versus magnetic field for selected filling factors. **G**, Schematic illustration of a non-coplanar spin texture. The finite scalar spin chirality generates an emergent magnetic field, giving rise to a topological Hall response in addition to the net magnetization. **H**, Decomposition of $R_{xy}$ at $\nu = -0.93$ into linear ($R_{LH}$), anomalous ($R_{AH}$), and topological ($R_{TH}$) Hall contributions, using RMCD as a proxy for magnetization. **I**, Extracted maximum topological Hall resistance $R_{TH}$ as a function of $\nu$, showing a pronounced peak near $\nu \approx -0.93$. Inset, schematic of a non-coplanar spin texture subtending a solid angle $\Omega$ and imparting a Berry phase.

# Supplementary Materials for

## Van Hove Singularity Driven Topological Magnetism in Twisted $MoTe_2$

Heonjoon Park[1], Julian Stewart[2], Xiao-Wei Zhang[2], Taige Wang[3,4], Canxun Zhang[5], Evgeny Redekop[5], Chaowei Hu[1], Jiaqi Cai[1], Weijie Li[1], Eric Anderson[1], Takashi Taniguchi[6], Kenji Watanabe[7], Jiun-Haw Chu[1], David Cobden[1], Andrea Young[5], Liang Fu[4], Ting Cao[2], Di Xiao[1,2], Xiaodong Xu[1,2*]

Corresponding author: xuxd@uw.edu

**The PDF file includes:**

## Materials and Methods

Section 1.1 Sample fabrication

The devices were assembled using a polymer-based dry-transfer method to construct a dual-gated heterostructure. Few-layer hexagonal boron nitride (hBN) and graphite flakes were mechanically exfoliated onto oxidized silicon substrates and screened by optical contrast. Thicknesses were verified by atomic force microscopy (AFM). A local bottom-gate stack was first prepared. Using a poly(bisphenol A carbonate) film supported on a polydimethylsiloxane (PDMS) stamp attached to a glass slide, we sequentially transferred an hBN flake followed by a graphite flake. The assembled stack was released onto a 90 nm $SiO_2$/Si wafer under controlled heating to ensure uniform adhesion. Electron-beam lithography was then employed to define a Hall-bar geometry on the bottom graphite gate. After development, Ti/Pt (2 nm/5 nm) electrodes were deposited by electron-beam evaporation. Contact-mode AFM scanning was subsequently used to remove resist residue and improve surface cleanliness.

$MoTe_2$ flakes were exfoliated inside a nitrogen-filled glovebox with $O_2$ and $H_2O$ concentrations maintained below 0.1 ppm. A selected monolayer flake was cut into two sections using an AFM tip to enable controlled twist assembly. For the contact-gate dielectric, a thin hBN layer was first picked up with the polymer stamp, followed by sequential stacking of the two $MoTe_2$ halves. After picking up the first half, the stage was rotated by a designated twist angle before picking up the second half, forming a twisted homobilayer structure. The completed stack was then released onto the pre-patterned bottom-gate substrate.

Post-transfer, the heterostructure was scanned in contact-mode AFM to expel interfacial contaminants and minimize trapped bubbles or strain-induced wrinkles. A patterned Ti/Pt (2 nm/5 nm) contact gate was subsequently deposited on top of the Hall-bar region to locally tune carrier density and reduce contact resistance. The device surface was again AFM-cleaned following metal deposition. Finally, a top-gate assembly consisting of hBN/graphite/hBN was transferred to complete the dual-gate architecture. Electrical connections to the buried Pt structures were defined by evaporating Cr/Au (7 nm/120 nm) leads extending to large bonding pads suitable for wire bonding.

Section 1.2 Transport measurements

Electrical transport measurements were performed in a Bluefors dilution refrigerator with a 13.5 T out-of-plane magnet with a base phonon temperature of 20 mK. Longitudinal resistance $R_{xx}$ and Hall resistance $R_{xy}$ were measured using a standard low-frequency lock-in technique in a four-terminal configuration with a low excitation current of 0.2–1 nA. To reduce mixing between longitudinal and Hall voltage signals caused by device asymmetry, resistance data were symmetrized with respect to magnetic field for $R_{xx}$ and antisymmetrized for $R_{xy}$.

Section 1.3 Optical measurements

Magneto-optical measurements were performed using reflective magnetic circular dichroism (RMCD) microscopy to probe out-of-plane magnetization in the moiré minibands. The measurements were carried out in a cryogenic optical system (attoDry2100) integrated with a superconducting magnet and electrical feedthroughs that enabled simultaneous optical probing and electrostatic gating of the device.

A broadband light source from a supercontinuum laser (NKT Photonics) was spectrally filtered to select the resonant photon energy and focused onto the device using a high-numerical-aperture objective in reflection geometry. Circular polarization modulation was generated using a photoelastic modulator (Hinds Instruments), allowing rapid switching between left- and right-circularly polarized light. The reflected signal from the sample was collected by the same objective and directed to an InGaAs avalanche photodetector, where the difference in reflectance between the two polarizations was measured using lock-in detection (SR830). This differential signal corresponds to the RMCD response and provides a direct probe of the sample's out-of-plane magnetization.

Section 1.4 NanoSQUID measurements

Local magnetometry measurements were carried out using a superconducting quantum interference device integrated at the apex of a pulled quartz pipette, commonly referred to as a nanoSQUID-on-tip (nSOT). The fabrication procedure followed established protocols reported previously(*64-66*). A quartz capillary with an inner diameter of 0.5 mm was thermally drawn into a sharp tip, yielding an apex size on the order of 150 nm. Electrical leads were defined by depositing Ti/Au (5 nm / 50 nm) onto the pipette using electron-beam evaporation at a rate of approximately 2 Å/s. A proximal shunt resistor, consisting of Ti/Au (8 nm / 15 nm), was subsequently patterned within roughly 500 μm of the tip apex, resulting in a resistance of about 3 Ω. To ensure robust electrical contact at cryogenic temperatures, the contact pads were coated with a thick indium layer, which also improves mechanical coupling to the spring-loaded contacts in the probe holder.

The superconducting element was formed by depositing indium (critical temperature $T_c$ = 3.4 K) onto the tip using a custom thermal evaporation setup. Material was evaporated sequentially from three oblique angles to coat opposing sides of the apex at approximately 110° relative to the tip axis, followed by a normal-incidence deposition. During this process, the probe assembly was thermally anchored to a cryogenic stage and enclosed within a liquid nitrogen jacketing, maintaining the temperature below 10 K. Prior to each deposition step, the system was allowed to equilibrate for 5–15 minutes in helium exchange gas at a pressure of $5 \times 10^{-3}$ mbar. Typical indium thicknesses ranged from 30 to 60 nm for angled depositions and 20 to 50 nm for the axial deposition, with a growth rate of ~0.1 nm/s, producing continuous superconducting films with fine grain structure at the tip apex.

Measurements were performed in a helium cryostat with a base temperature of 1.7 K. The magnetic flux threading the nanoSQUID was detected using a quasi-voltage-biased readout, incorporating a series SQUID array amplifier (SSAA) together with a compensation circuit(*67*). Sensor calibration was achieved by recording the voltage noise spectrum of the SSAA and converting it to an equivalent magnetic noise using the magnetic field-to-voltage transfer function, which was determined from the device response to a calibrated 20 μT magnetic field step. From this procedure, we obtained a magnetic field sensitivity in the range of 1–10 nT/√Hz. Given an effective pickup loop diameter of approximately 200 nm, this corresponds to a flux sensitivity of about 20–200 $n\Phi_0/\sqrt{Hz}$.

Section 1.5. Estimation of carrier density, electric field, and moiré filling factor

The carrier density $n$ and perpendicular electric displacement field $D/\varepsilon_0$ were determined from the applied top- and bottom-gate voltages $V_{\mathrm{tg}}$ and $V_{\mathrm{bg}}$ using a parallel-plate capacitance model. The relations are

$$n = (C_{tg}V_{tg} + C_{bg}V_{bg})/e - n_{\mathrm{offset}},$$
$$D/\varepsilon_0 = (C_{\mathrm{tg}}V_{\mathrm{tg}} - C_{\mathrm{bg}}V_{\mathrm{bg}})/2\varepsilon_0 - D_{\mathrm{offset}}/\varepsilon_0,$$

where $e$ is the elementary charge and $\varepsilon_0$ is the vacuum permittivity. $C_{\mathrm{tg}}$ and $C_{\mathrm{bg}}$ denote the geometric capacitances of the top and bottom gates, respectively. These capacitances were estimated from the thickness of the hexagonal boron nitride dielectric layers measured by atomic force microscopy. The offset parameters $n_{\mathrm{offset}}$ and $D_{\mathrm{offset}}$ account for residual doping and built-in electric fields arising from fabrication asymmetries. $D_{\mathrm{offset}}/\varepsilon_0$ was found to be approximately 40 mV/nm for the 4.8° device. The value of $n_{\mathrm{offset}}$ was determined from the charge neutrality point in the dual-gate resistance maps.

To refine the capacitance estimates, we used high-field magnetotransport measurements to construct Landau fan diagrams. The slopes of the Landau level trajectories in the $n$-$B$ plane provide an independent calibration of the carrier density, enabling accurate determination of the effective gate capacitances. The capacitance values extracted from the Landau fan analysis were in good agreement with those obtained from the geometric capacitance model.

The moiré filling factor $\nu$, defined as the number of holes per moiré unit cell, was obtained by converting the carrier density using the moiré unit-cell area. In practice, the assignment of $\nu$ was verified by tracking the Landau fan features toward zero magnetic field and aligning them with the strongly developed correlated insulating states observed at fractional fillings. In particular, correlated insulating states appearing at $\nu = -1/3$, $-1/4$, $-1/6$, and $-1/7$ served as reliable markers for calibrating the filling factor across the dual-gated phase diagram (Fig. S5). This procedure ensured a consistent mapping between gate voltages, carrier density, and moiré band filling.

## Supplementary Text

Section 2.1. Lattice relaxations and band structure calculations

We employ machine learning force fields (MLFFs) to perform the moiré lattice relaxation using the LAMMPS package(*68*). The MLFFs are parameterized using the deep potential molecular dynamics (DPMD) method(*69, 70*). The training dataset is generated from ab initio molecular dynamics (AIMD) simulations for a 6° $tMoTe_2$ at 500 K using the VASP package(*71*). More details regarding the parameterization of the MLFFs can be found in Ref(*72*).

To efficiently compute the DOS under a displacement field, we fit a continuum model to the large-scale DFT moiré band structures at twist angles of $5.09^{\circ}$ and $4.41^{\circ}$. The DFT band structures are calculated using the SIESTA package(*73*). The continuum Hamiltonian for the K valley is written as

$$H_K^{\uparrow} = \begin{pmatrix} -\frac{\hbar^2(\boldsymbol{k}-\boldsymbol{K}_{\mathrm{b}})^2}{2m^*} + \Delta_{\mathrm{b}}(\boldsymbol{r}) + \frac{V_D}{2} & \Delta_T(\boldsymbol{r}) \\ \Delta_{\mathrm{T}}^{\dagger}(\boldsymbol{r}) & -\frac{\hbar^2(\boldsymbol{k}-\boldsymbol{K}_{\mathrm{t}})^2}{2m^*} + \Delta_{\mathrm{t}}(\boldsymbol{r}) - \frac{V_D}{2} \end{pmatrix}.$$

Here, the intralayer potentials $\Delta_{\mathrm{b/t}}(\boldsymbol{r})$ and the interlayer tunneling $\Delta_T(\boldsymbol{r})$ are expanded within the first-harmonic approximation,

$$\Delta_{\mathrm{b/t}}(\boldsymbol{r}) = 2v \sum_{j=1,3,5} cos\,(\boldsymbol{G}_j \cdot \boldsymbol{r} \pm \psi),$$
$$\Delta_T(\boldsymbol{r}) = w(1 + e^{-i\boldsymbol{G}_2\cdot\boldsymbol{r}} + e^{-i\boldsymbol{G}_3\cdot\boldsymbol{r}}).$$

The parameter $V_D$ describes the potential difference between the two layers induced by the displacement field. The effective mass is taken as $m^*$=0.672 $m_e$ .

The three parameters $v$, $\psi$, and $w$ are determined by fitting the first two moiré bands for each valley at zero displacement field, as shown in Fig. S1a. The fitted values are ( $v = 17.8$ meV, $\psi = 134^{\circ}, w = -18.4$ meV) . We note that the fitted parameters change negligibly between twist angles $5.09^{\circ}$ and $4.41^{\circ}$, therefore, the twist-angle dependence of these parameters is neglected.

To determine the displacement-field dependence of $V_D$ , we first fit $V_D$ at $D/\varepsilon_0 = 100$ mV/nm while keeping the other parameters fixed. Figure S1b compares the continuum model with the DFT results. At this displacement field, the fitted value is $V_D = 7.6$ meV . For other displacement fields, $V_D$ is obtained by linear scaling. The continuum model is then used to calculate the band structure and DOS up to 600 mV/nm. A dense $240 \times 240$ k-grid is used to sample the moiré Brillouin zone in the DOS calculations.

Section 2.2. Hartree-Fock calculations

We study interaction effects using self-consistent Hartree-Fock (HF) calculations based on the continuum model introduced above (*74, 75*). To reduce the computational cost, the Coulomb interaction is projected onto the highest two moiré valence bands in each valley. Keeping two bands per valley is essential because the gap between the first two moiré valence bands becomes small at finite displacement field, and mixing with the second moiré band can change the Chern number. The interacting Hamiltonian is written as

$$H = H_0 + H_C + H_Z, \qquad H_C = \frac{1}{2A}\sum_{\mathbf{q}} V_C(\mathbf{q}) :\rho(\mathbf{q})\rho(-\mathbf{q}):,$$

where $H_0$ is the continuum Hamiltonian in Section 2.1, $A$ is the sample area, and $V_C(\mathbf{q}) = e^2 \tanh(q d_g)/(2\varepsilon q)$ is the dual-gate screened Coulomb interaction. Here $d_g = 30\,\mathrm{nm}$ is the sample-gate distance and the dielectric constant is taken as $\varepsilon = 50$. A $30 \times 30$ $k$-grid is used to sample the moiré Brillouin zone in self-consistent calculations.

In the calculations shown in the main text, the twist angle is taken to be $4.8^{\circ}$, displacement field is fixed at $D/\epsilon_0 = 300\,\mathrm{mV/nm}$ and the electron filling is $\nu = -0.995$. We retain both valley-diagonal and valley-off-diagonal components of the one-body density matrix, thereby allowing both valley-polarized and valley-coherent solutions. For the valley-off-diagonal channel, we adopt a single-$\mathbf{Q}$ ansatz, so that the intervalley coherence takes the form

$$\langle \psi^{\dagger}_{n,\tau,\mathbf{k}} \psi_{n',-\tau,\mathbf{k}+\mathbf{Q}} \rangle.$$

We compare all $\mathbf{Q}$'s and find that $\mathbf{Q} = \boldsymbol{\kappa}_-$ has the lowest HF energy in the parameter regime shown in the main text. We further include a Zeeman field defined as the energy difference between the two valleys,

$$H_Z = -\frac{E_Z}{2}\sum_{n,\mathbf{k}} \left( c^{\dagger}_{n,K,\mathbf{k}} c_{n,K,\mathbf{k}} - c^{\dagger}_{n,K',\mathbf{k}} c_{n,K',\mathbf{k}} \right),$$

and neglect the orbital effect of the magnetic field.

The Hall conductance, valley $U(1)$-breaking order parameter $U_{\mathrm{v}}(1)$, and valley polarization $P_{\mathrm{v}}$ shown in the main text are extracted from the self-consistent HF bands as functions of $E_Z$. The valley $U(1)$-breaking order parameter is characterized by the magnitude of the valley-off-diagonal HF density matrix, while the valley polarization is defined from the occupation difference between the two valleys. The Hall conductance is computed from the Berry curvature of the occupied part of the HF bands and reported in units of $e^2/h$,

$$\sigma_{xy} = \frac{e^2}{h}\frac{1}{2\pi}\sum_{n}\int_{\mathrm{occ}} d^2k\, \Omega_n(\mathbf{k}),$$

where the integral is taken over the states below the HF chemical potential.

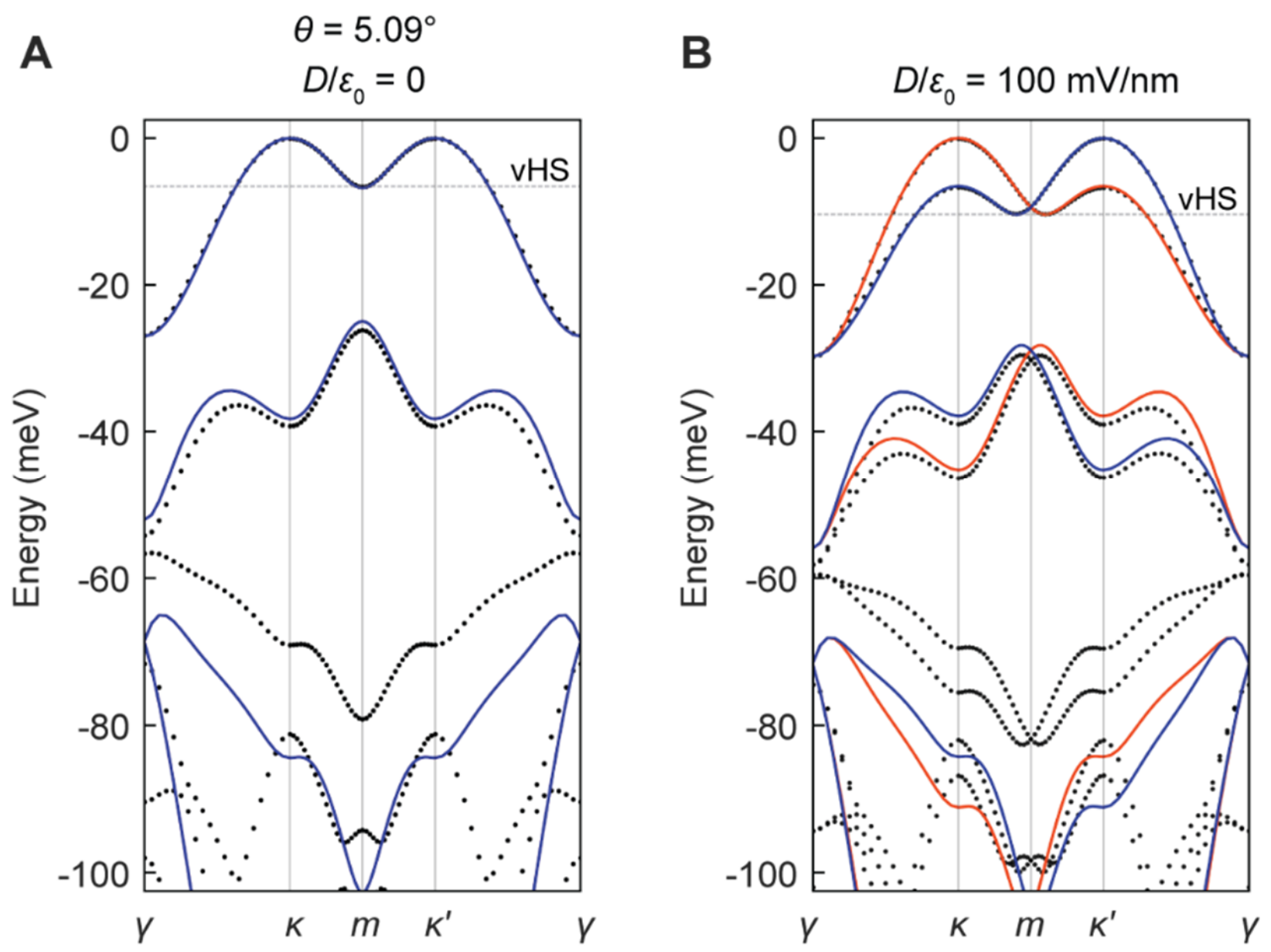

**Fig. S1. DFT calculations and continuum model.** (**A**) Band structures obtained from the fitted continuum model (blue and red lines) and DFT calculations (black dots) at zero displacement field and $\mathbf{5.09}^{\circ}$. Dotted gray line indicates the van Hove singularity (vHS) saddle point in the first valence band. (B) Band structures from the fitted continuum model (blue and red lines) and DFT calculations (black dots) at $D/\varepsilon_0$ = 100 mV/nm and $\mathbf{5.09}^{\circ}$.

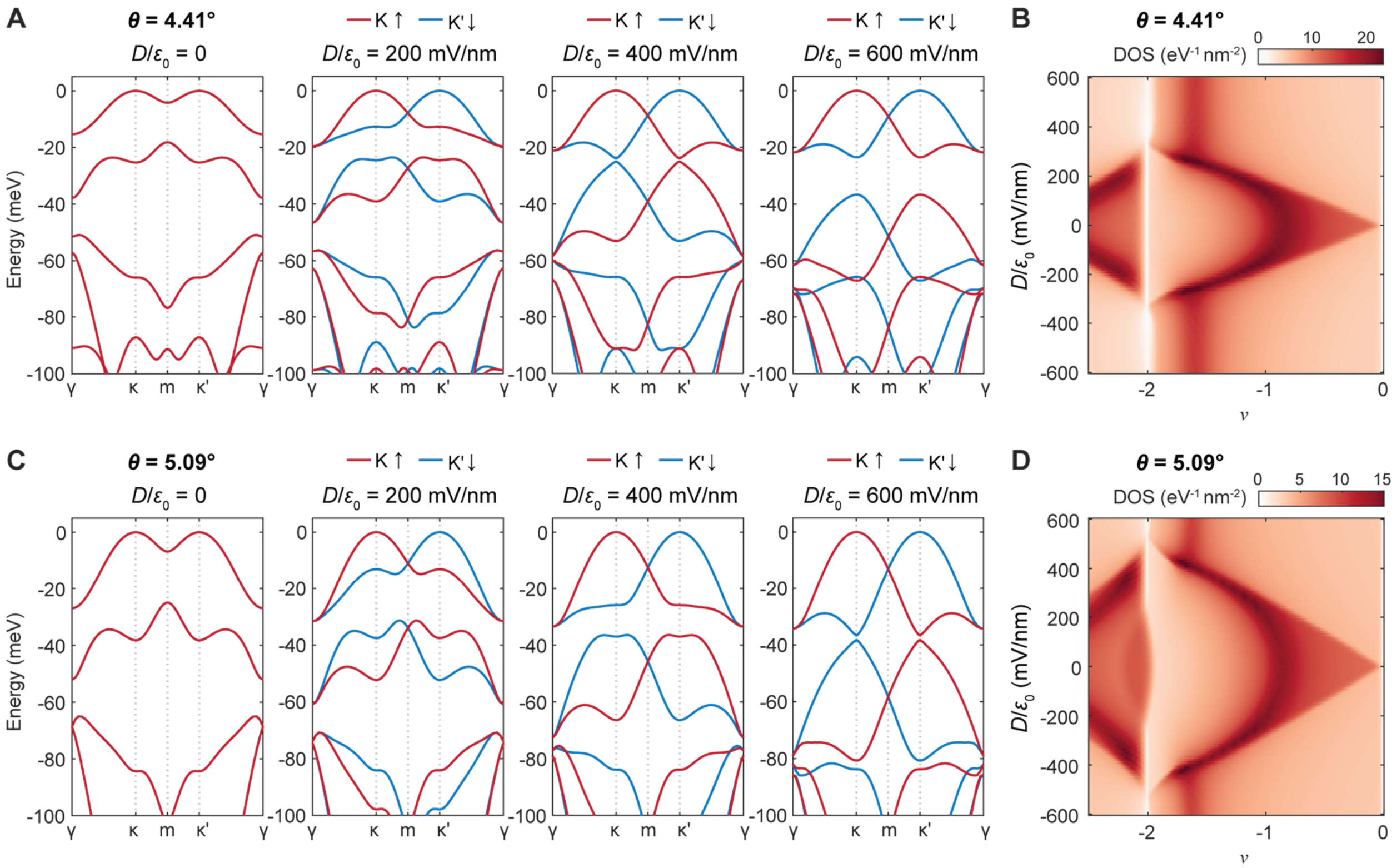

**Fig. S2. Continuum model calculations at different twist angles and electric fields.** (**A**) Electronic band structure of 4.41° twisted $MoTe_2$ homobilayer at different electric fields ($D/\varepsilon_0$). Red (blue) lines represent spin up K (spin down K') bands. (**B**) Calculated density of states as a function of filling factor ($v$) and electric field ($D/\varepsilon_0$). The van Hove singularity arc around $v \approx -1$ moves toward higher displacement field as filling factor increases, and eventually separates into three different branches. (**C**), (**D**), similar calculations for 5.09° twisted $MoTe_2$. The bands are more dispersive due to larger twist angle.

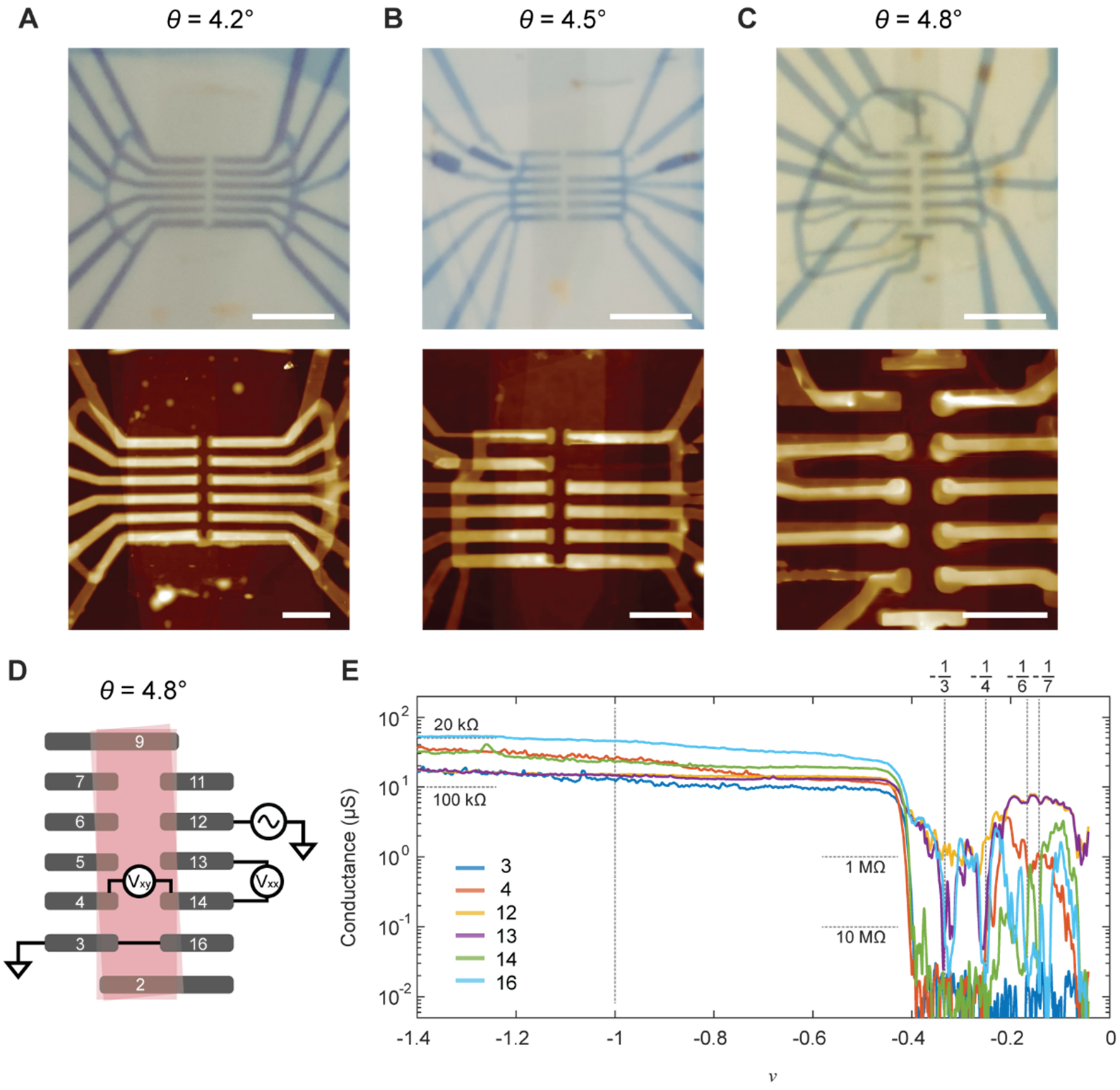

**Fig. S3. Device images and contact characterization.** (**A-C**) Optical microscope images (top) and AFM images (bottom) of twisted $MoTe_2$ devices with twist angles of 4.2° (**A**), 4.5° (**B**), and 4.8° (**C**). Scale bars correspond to 5 μm for the optical images and 2 μm for the AFM images. (**D**) Measurement configuration used for the 4.8° device, indicating the contact geometry employed to extract longitudinal and Hall resistances. (**E**) Two-terminal contact conductance as a function of filling factor $\nu$ measured at $T = 20$ mK for the contacts used in transport measurements. Current was sourced through a single contact while all remaining contacts were grounded. The contacts exhibit conductances corresponding to resistances below 100 kΩ over most of the density range, increasing only when the device approaches strongly insulating regimes associated with fractional correlated states.

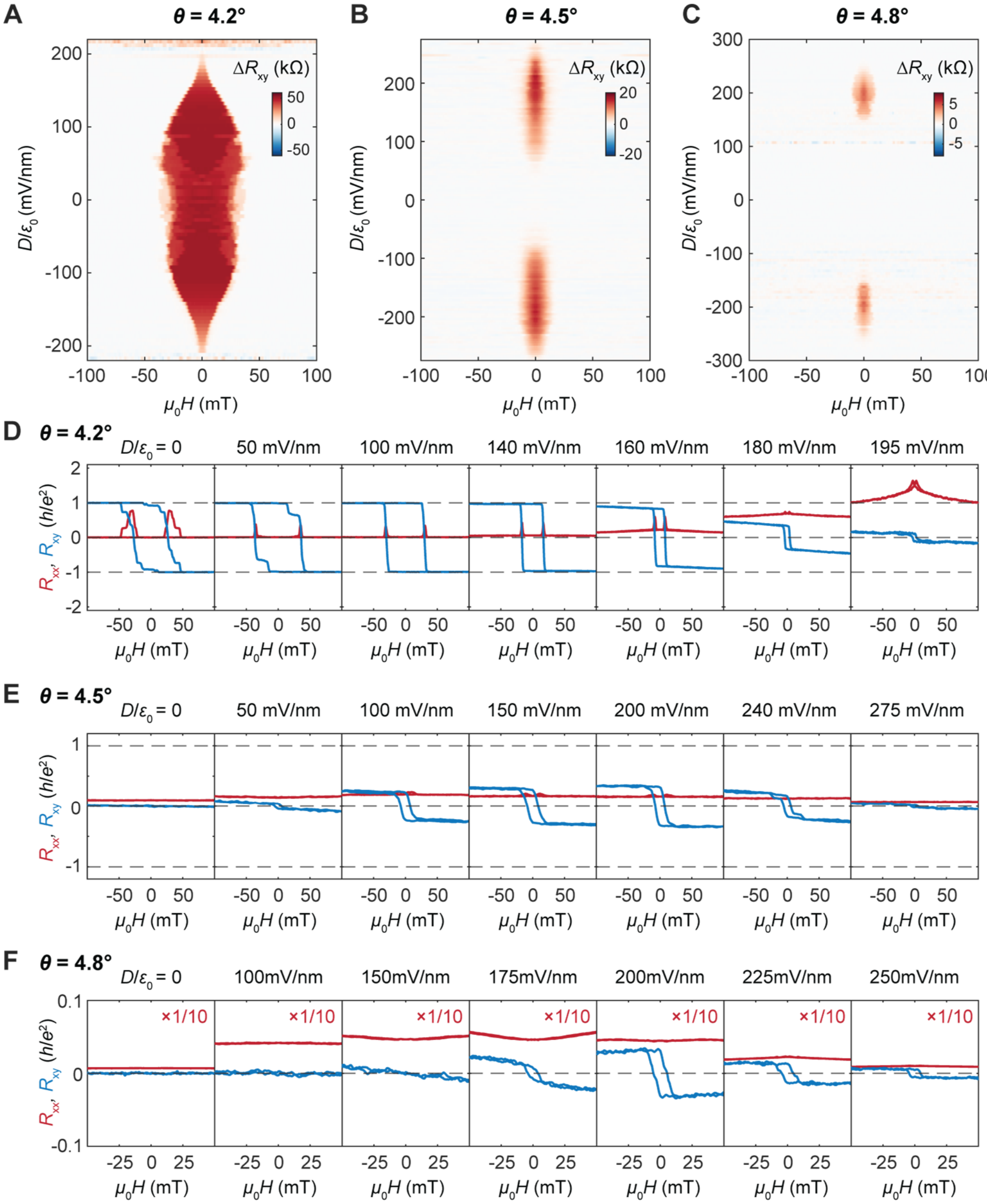

**Fig. S4. Twist-angle and electric-field dependence of the anomalous Hall effect.** (**A-C**) Hysteretic component of the Hall resistance, $\Delta R_{xy}$, plotted versus perpendicular magnetic field ($\mu_0 H$) and electric displacement field ($D/\varepsilon_0$) for devices with twist angles $\theta$ = 4.2° (**A**), 4.5° (**B**), and 4.8° (**C**). The ferromagnetic response near zero electric field weakens as the twist angle increases. (**D-F**) Longitudinal resistance $R_{xx}$ (red) and Hall resistance $R_{xy}$ (blue) measured during magnetic-field sweeps at selected electric fields for $\theta$ = 4.2° (**D**), 4.5° (**E**), and 4.8° (**F**). For $\theta$ = 4.2°, a quantum anomalous Hall state persists up to $D/\varepsilon_0 \approx 140$ mV/nm before being suppressed at larger displacement fields. At larger twist angles (4.5° and 4.8°), the anomalous Hall response is strongest at finite electric field rather than at $D/\varepsilon_0 = 0$.

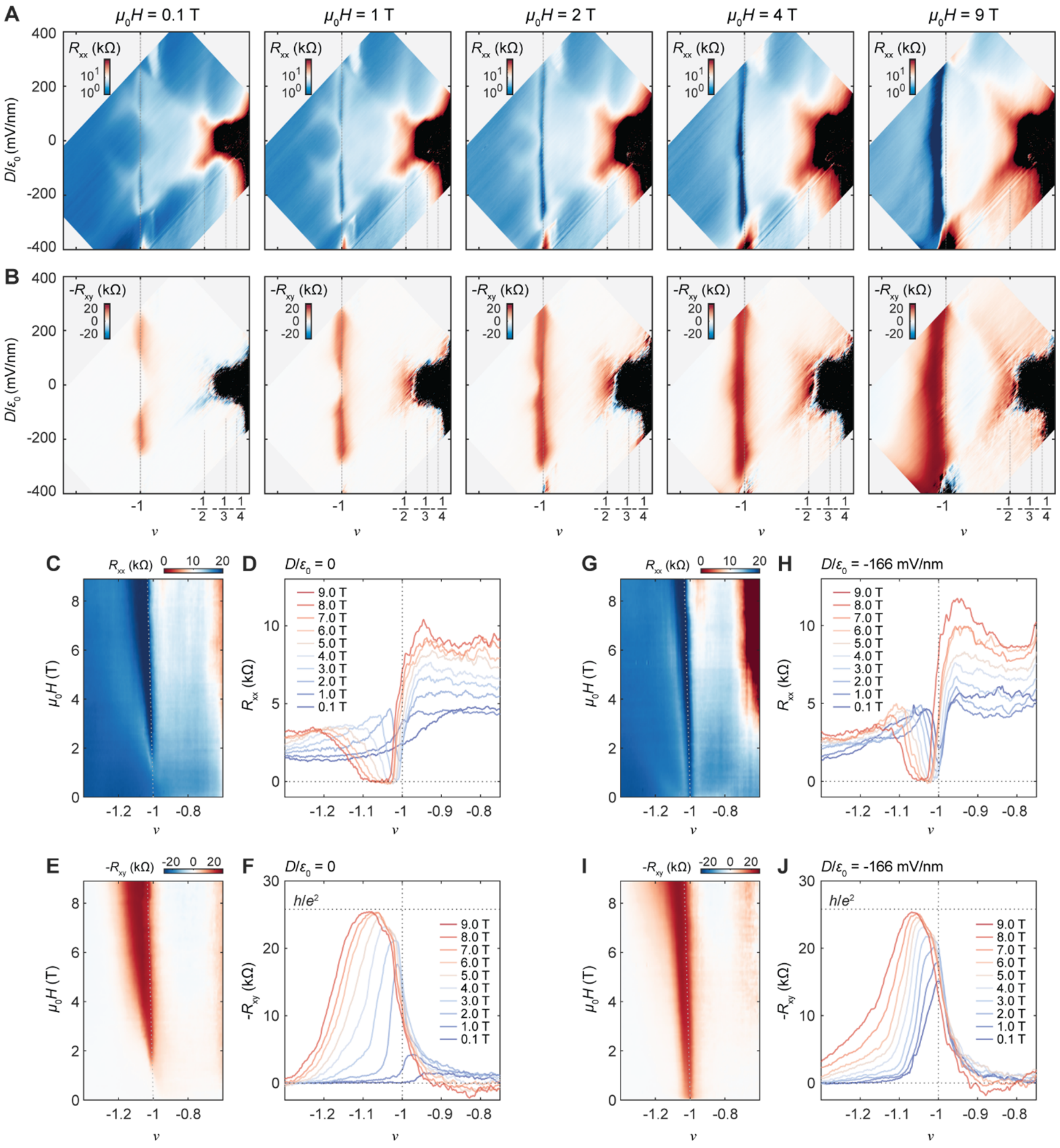

**Fig. S5. Emergent incipient Chern insulator state near the van Hove singularity in the 4.5° twisted device.** (**A-B**) Phase diagrams of longitudinal resistance $R_{xx}$ and Hall resistance $R_{xy}$ as a function of filling factor $v$ and electric displacement field $D/\varepsilon_0$, measured at different magnetic fields $\mu_0 H$. (**C-D**) $R_{xx}$ and (**E-F**) $R_{xy}$ as a function of $v$ and $\mu_0 H$ at $D/\varepsilon_0 = 0$. The progression follows the Streda formula (white dashed line) for $C = -1$ and the $R_{xx}$ vanishes while $R_{xy}$ is quantized at high magnetic field, instantiating the quantum anomalous Hall state. (**G-J**) Same as (**C-F**) but for $D/\varepsilon_0 = -166$ mV/nm.

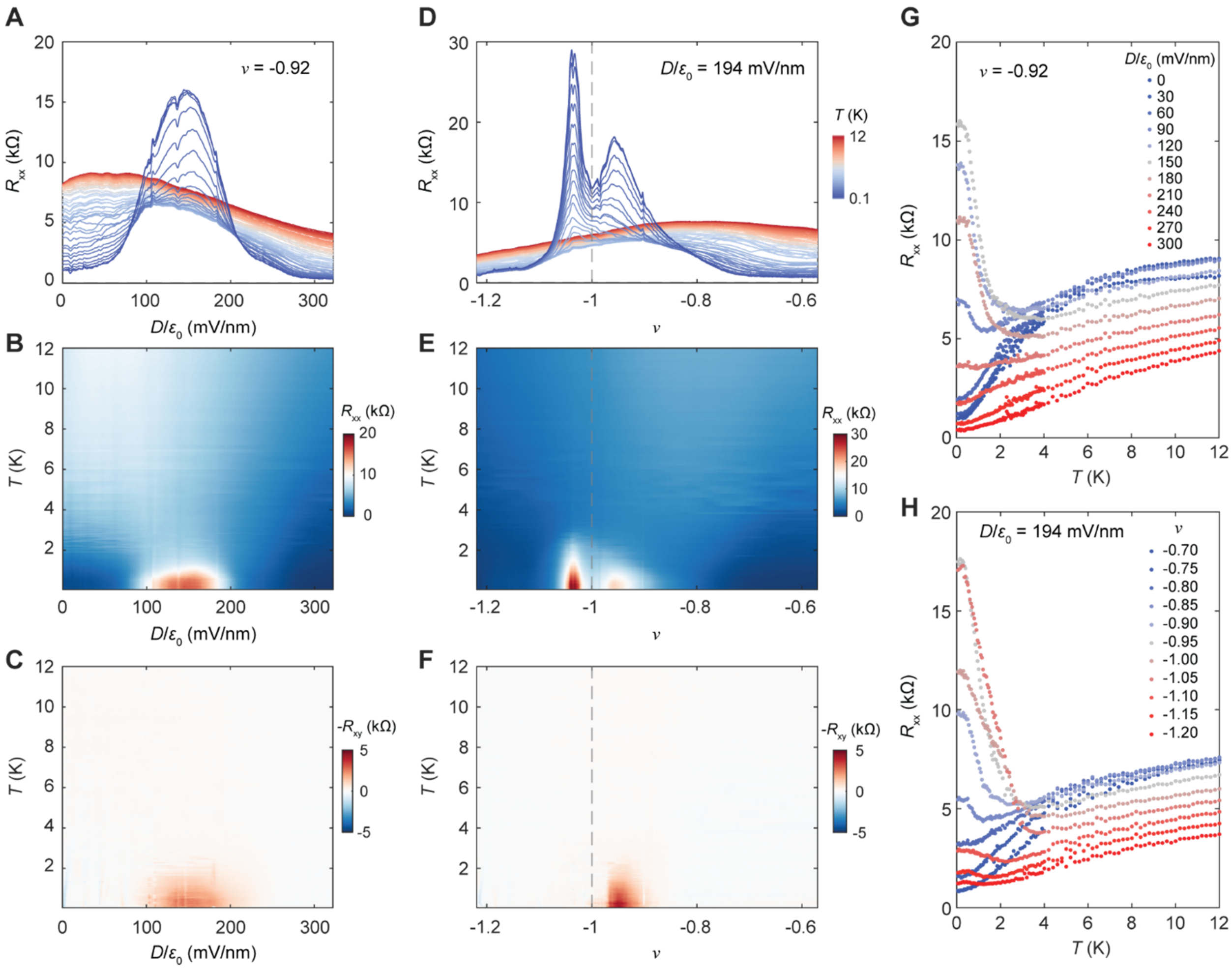

**Fig. S6. Anomalous Hall effect hotspot in the 4.8° twisted device.** (**A**–**B**) Longitudinal resistance $R_{xx}$ and (**C**) Hall resistance $R_{xy}$ as a function of electric displacement field $D/\varepsilon_0$ at different temperatures, measured at a fixed filling factor of $\nu = -0.92$. (**D**–**E**) $R_{xx}$ and (**F**) $R_{xy}$ as a function of filling factor at a fixed electric field of $D/\varepsilon_0 = 194$ mV/nm. (**G**) Temperature dependence of $R_{xx}$ at different electric fields for fixed filling $\nu = -0.92$, and (**H**) temperature dependence of $R_{xx}$ at different filling factors for a fixed electric field of $D/\varepsilon_0 = 194$ mV/nm. Near the anomalous Hall effect hotspot, both $R_{xx}$ and $R_{xy}$ increase with decreasing temperature and approach saturation.

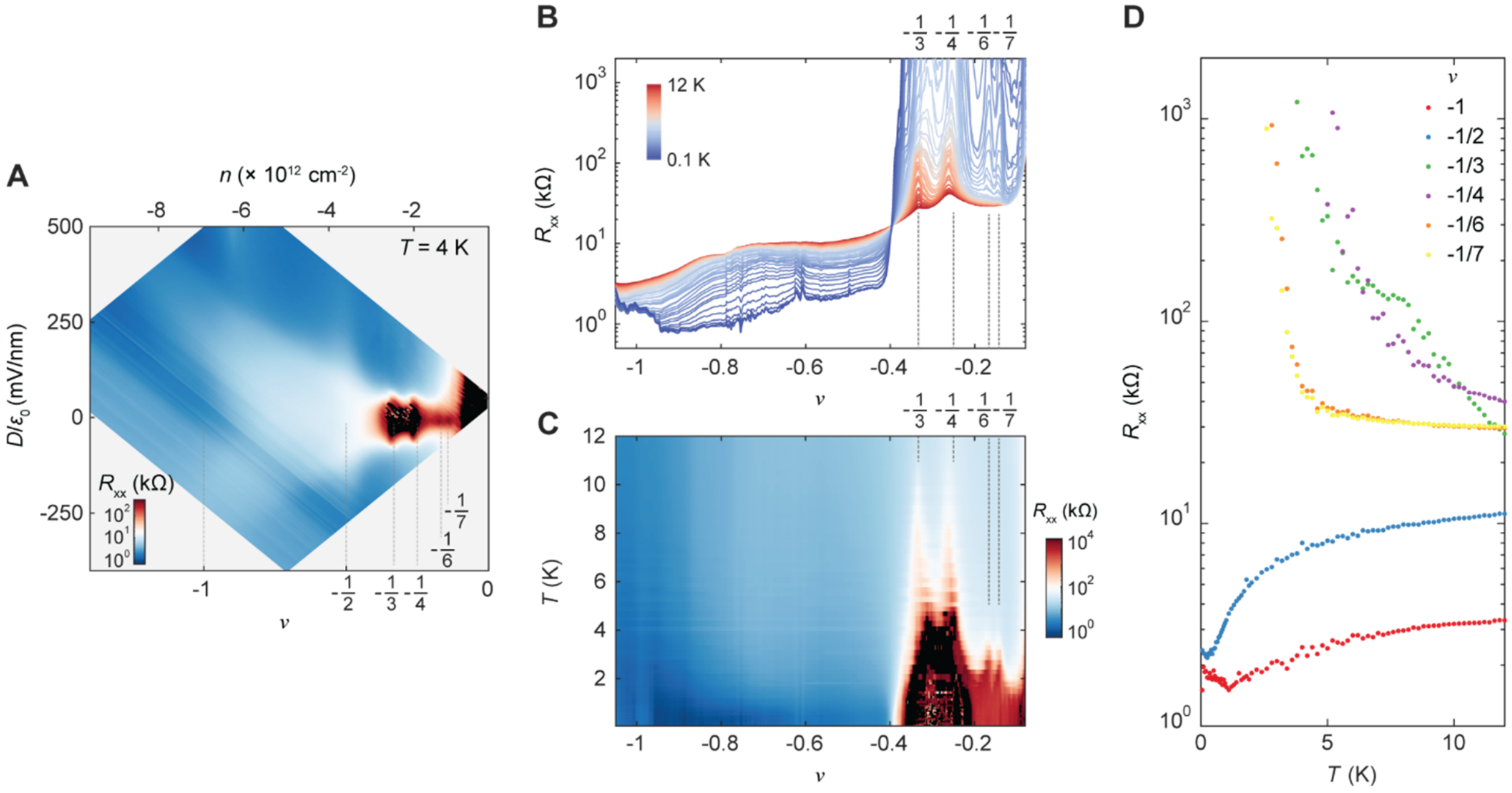

**Fig. S7. Correlated insulating states in the 4.8° twisted device.** (**A**) Longitudinal resistance $R_{xx}$ as a function of filling factor $\nu$ (top axis shows the corresponding carrier density $n$) and electric displacement field $D/\varepsilon_0$, measured at $T$ = 4 K. Distinct resistance enhancements appear near fractional fillings $\nu = -1/3, -1/4, -1/6$, and $-1/7$, indicating the emergence of correlated insulating states. (**B**) Line cuts of $R_{xx}$ versus filling factor at different temperatures, measured at $D/\varepsilon_0 = 0$. Vertical dashed lines mark the fractional fillings where insulating features develop. (**C**) Temperature dependence of $R_{xx}$ as a function of filling factor at $D/\varepsilon_0 = 0$. (**D**) $R_{xx}$ as a function of temperature at selected fillings, showing enhanced resistance and insulating behavior at the fractional fillings ($\nu = -1/3, -1/4, -1/6$, and $-1/7$) compared to the metallic behavior at $\nu$ = -1, -1/2.

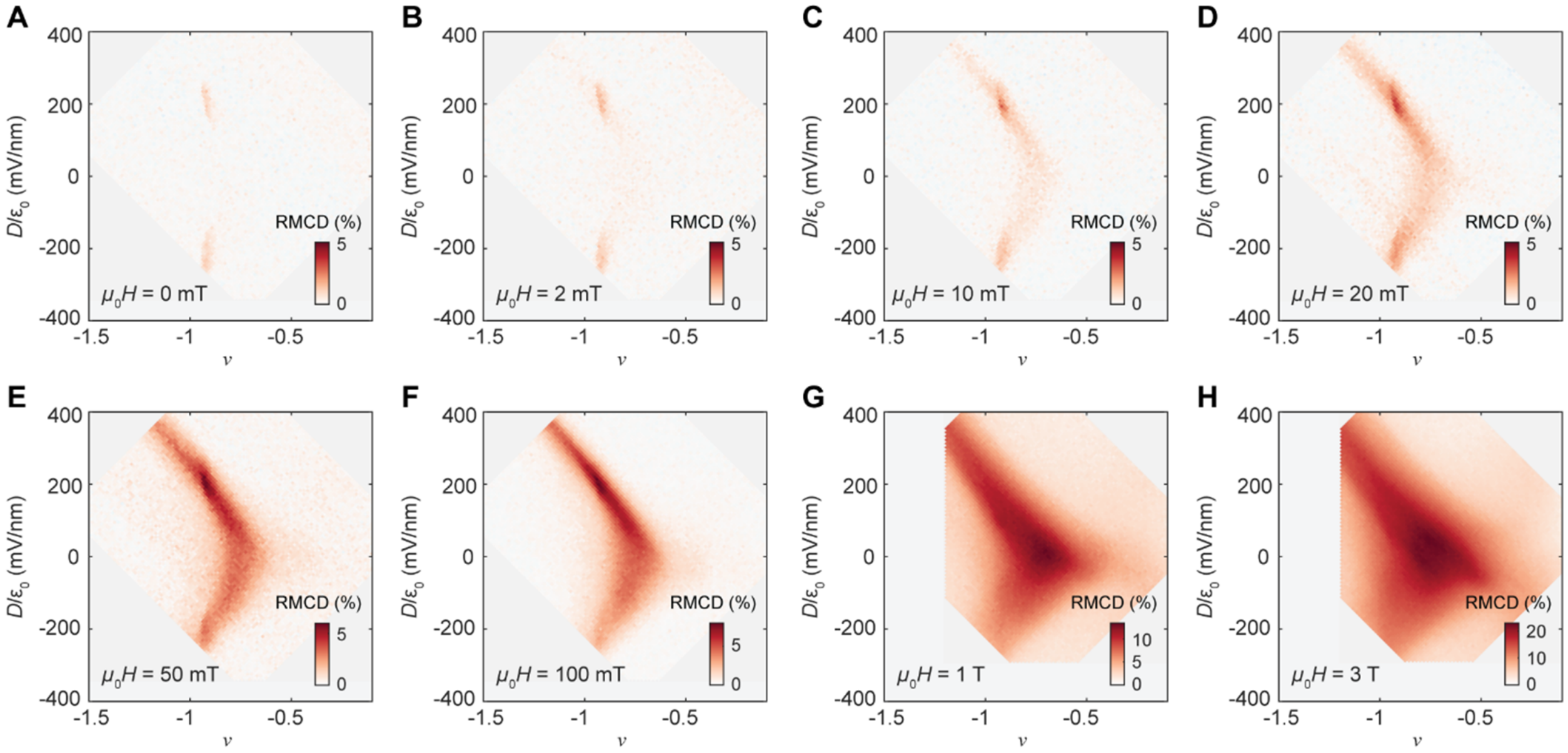

**Fig. S8. Phase diagram of magnetic circular dichroism at different magnetic fields.** (**A–H**) Phase diagrams of reflective magnetic circular dichroism (RMCD) as a function of filling factor $v$ and electric field $D/\varepsilon_0$, measured at different magnetic fields $\mu_0 H$ at $T$ = 1.6 K. At zero magnetic field, a localized RMCD signal appears near the anomalous Hall effect hotspot. With increasing magnetic field, the RMCD response strengthens and expands across the entire region near the van Hove singularity, indicating a progressively enhanced magnetic polarization of the electronic states.

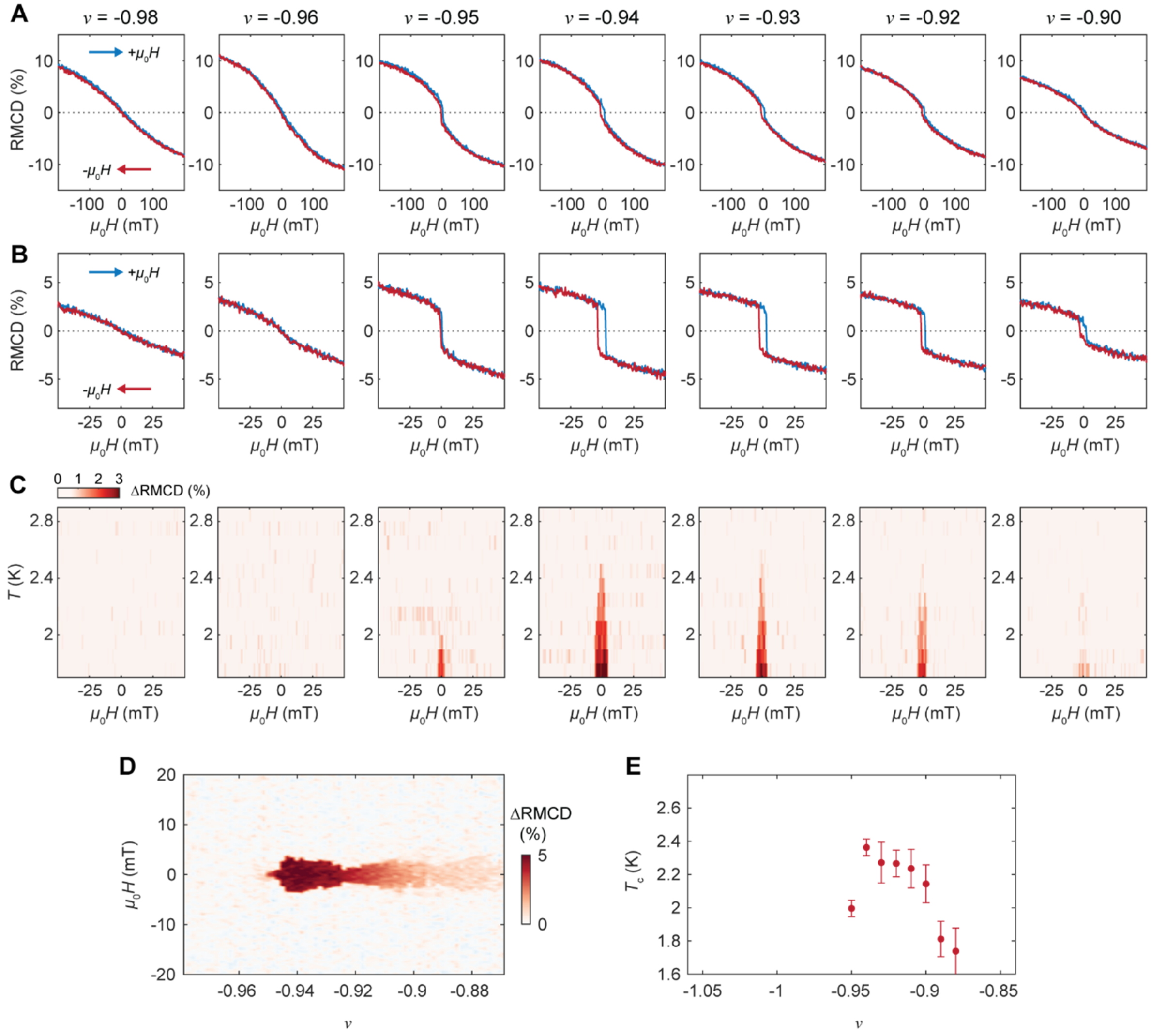

**Fig. S9. Filling factor-dependent RMCD and Curie temperature extraction.** (**A**) Reflective magnetic circular dichroism (RMCD) as a function of magnetic field for field sweeps in the positive (blue) and negative (red) directions at selected filling factors $\nu$, measured at fixed electric field $D/\varepsilon_0$ = 194 mV/nm. (**B**) Expanded view of the low-field RMCD traces in (**A**), highlighting the hysteresis that develops upon doping away from $\nu$ = -1. (**C**) Temperature dependence of the hysteretic component of the RMCD (ΔRMCD) plotted versus magnetic field for the same filling factors. (**D**) ΔRMCD as a function of filling factor and magnetic field, showing that the hysteretic magnetic response is concentrated in a narrow density window centered near the anomalous Hall effect hotspot. (**E**) Filling dependence of the extracted Curie temperature $T_c$ obtained from the temperature evolution in (**C**). For each filling factor, the hysteresis amplitude was defined as the maximum ΔRMCD near zero magnetic field within $|\mu_0 H| \leq 8$ mT after subtracting a background estimated from higher fields, 18 mT $\leq |\mu_0 H| \leq$ 40 mT. $T_c$ was defined as the highest temperature at which the net hysteretic signal remained above the background noise level, determined by linear interpolation between adjacent temperature points. Error bars represent the uncertainty associated with the background noise threshold used to identify the disappearance of hysteresis.

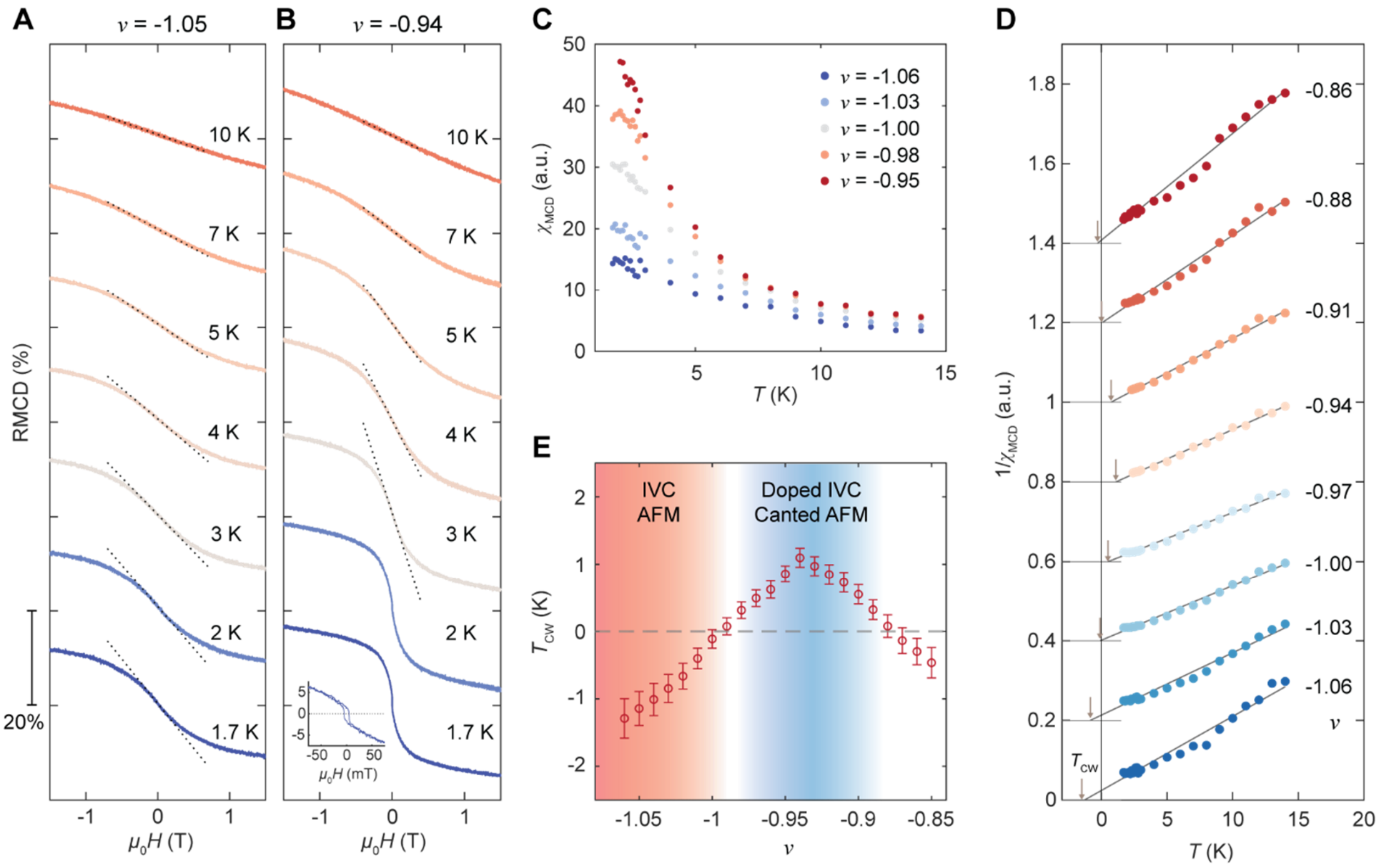

**Fig. S10. Magnetic susceptibility and Curie-Weiss analysis.** (**A**,**B**) Reflective magnetic circular dichroism (RMCD) as a function of magnetic field at filling factors $\nu = -1.05$ (**A**) and $\nu = -0.94$ (**B**) for selected temperatures. The magnetic susceptibility is estimated from the slope of the RMCD signal near zero field, defined as $\chi_{MCD} = (\boldsymbol{dRMCD/dH})|_{\boldsymbol{H\to 0}}$, following Ref.(*76, 77*). At $\nu = -0.94$, a small ferromagnetic hysteresis emerges at low temperatures, as shown in the inset of (**B**). The extraction of $\chi_{MCD}$ was limited to temperatures above the onset of hysteresis to avoid contributions from ferromagnetic ordering. (**C**) Temperature dependence of $\chi_{MCD}$ for several filling factors near $\nu = -1$. (**D**) Inverse susceptibility $1/\chi_{MCD}$ plotted as a function of temperature. The data are fit to the Curie–Weiss form $\boldsymbol{\chi = C/(T - T_{CW})}$. The grey arrows indicate the intercepts of the linear fits, corresponding to the Curie–Weiss temperature $T_{CW}$ for each filling. (**E**) Extracted Curie–Weiss temperature $T_{CW}$ as a function of filling factor. Error bars represent uncertainties from the linear fits in (**D**). $T_{CW}$ remains negative up to $\nu \approx -1$, indicating dominant antiferromagnetic interactions consistent with a 120° Néel order in the intervalley coherent (IVC) state. With further hole doping, $T_{CW}$ becomes positive and reaches a maximum of ~1 K near $\nu \approx -0.94$, suggesting the onset of ferromagnetic correlations due to canting of the 120°-ordered spins.

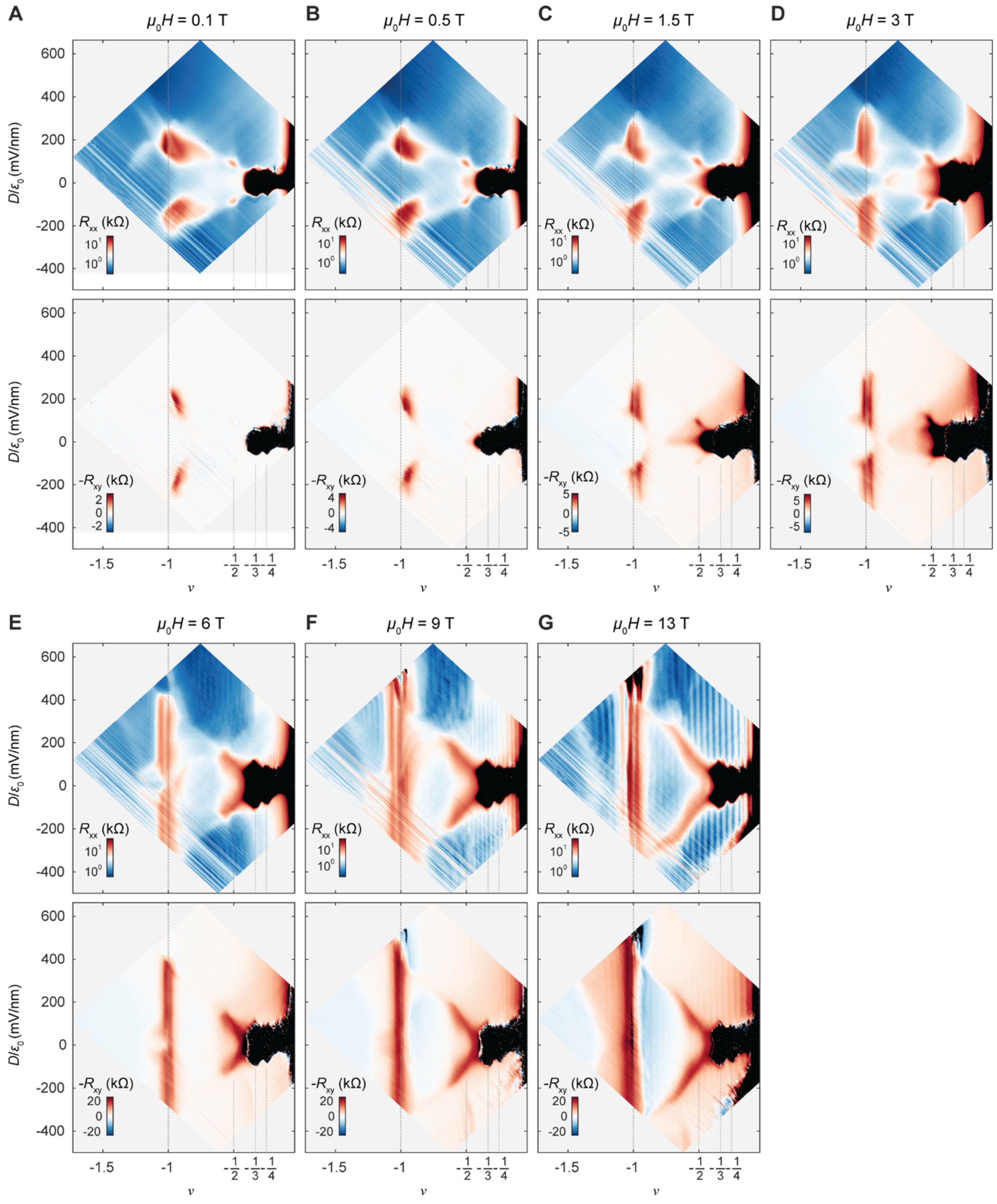

**Fig. S11. Transport phase diagram as a function of magnetic field.** (**A–G**) $R_{xx}$ (top row) and $R_{xy}$ (bottom row) plotted as a function of $\nu$ and $D/\varepsilon_0$, measured at different $\mu_0 H$.

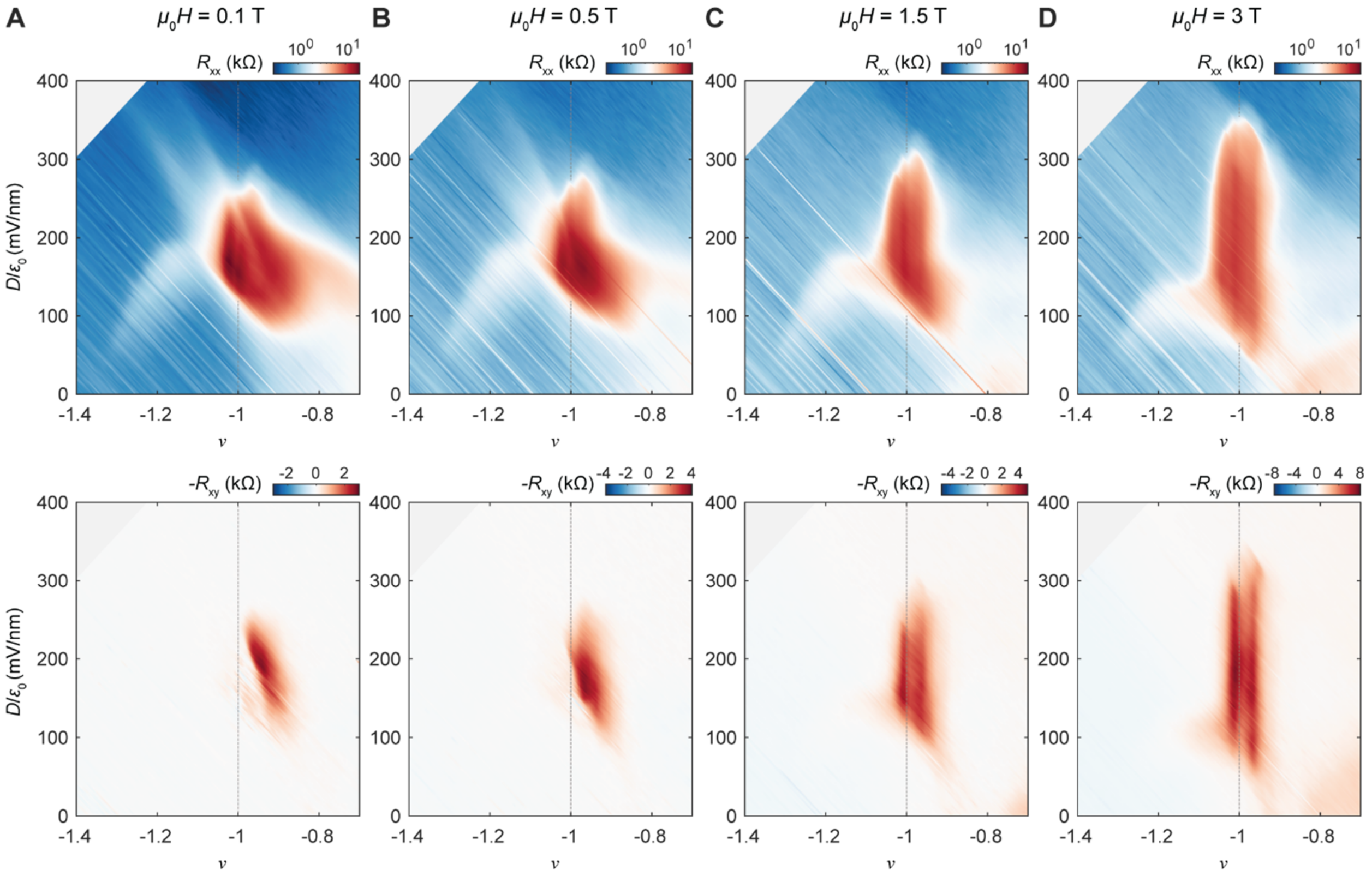

**Fig. S12. Zoomed-in transport phase diagram at low magnetic field near *ν* = -1.** (**A–D**) Top row: $R_{xx}$ as a function of filling factor *ν* and electric field $D/\varepsilon_0$, measured at $\mu_0 H$ = 0.1, 0.5, 1.5, and 3 T, respectively. Bottom row: corresponding -$R_{xy}$ maps measured under the same conditions. As magnetic field increases, the high-resistance region in $R_{xx}$ extends over a wider range of electric field, while the corresponding anomalous Hall response in $R_{xy}$ strengthens and broadens, eventually resulting in a field-stabilized Chern-insulator.

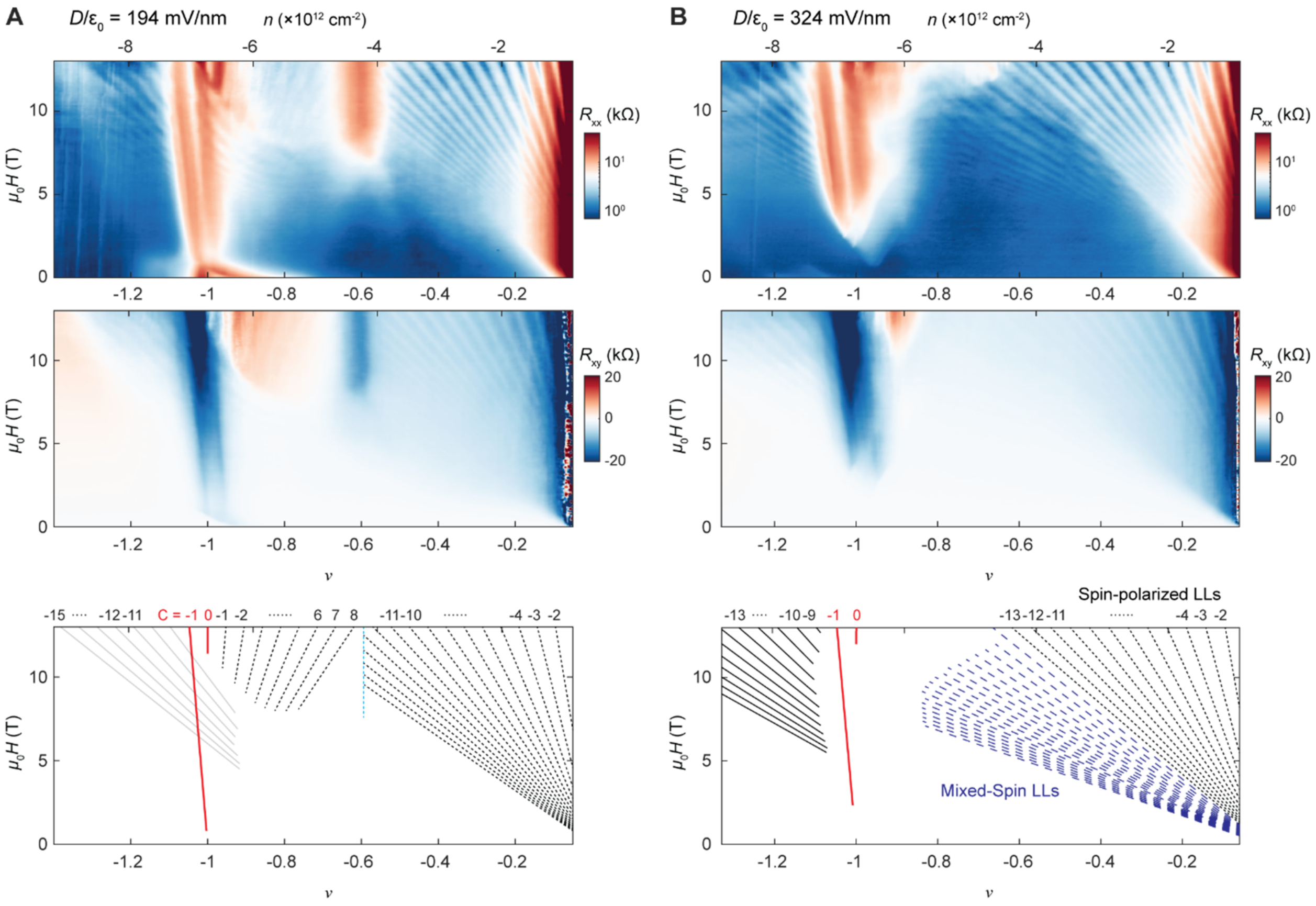

**Fig. S13. Landau fan at selected electric fields.** (**A**, **B**) Top row: $R_{xx}$ as a function of filling factor $\nu$ and magnetic field $\mu_0 H$ measured at fixed electric field $D/\varepsilon_0$ = 194 mV/nm (**A**) and 324 mV/nm (**B**). Middle row: corresponding $R_{xy}$ maps measured under the same conditions. Bottom row: schematic Wannier diagrams, overlaid on the same $\nu$ and $\mu_0 H$ axes. The black dashed lines denote spin-polarized Landau levels, while the blue dashed lines indicate mixed-spin Landau levels. The red line marks the $C$ = -1 Zeeman-gapped Chern insulator state at $\nu$ = -1. Additional Landau level features may arise from smaller Fermi surface pockets undergoing Lifshitz transitions, although the underlying mechanism remains to be determined.

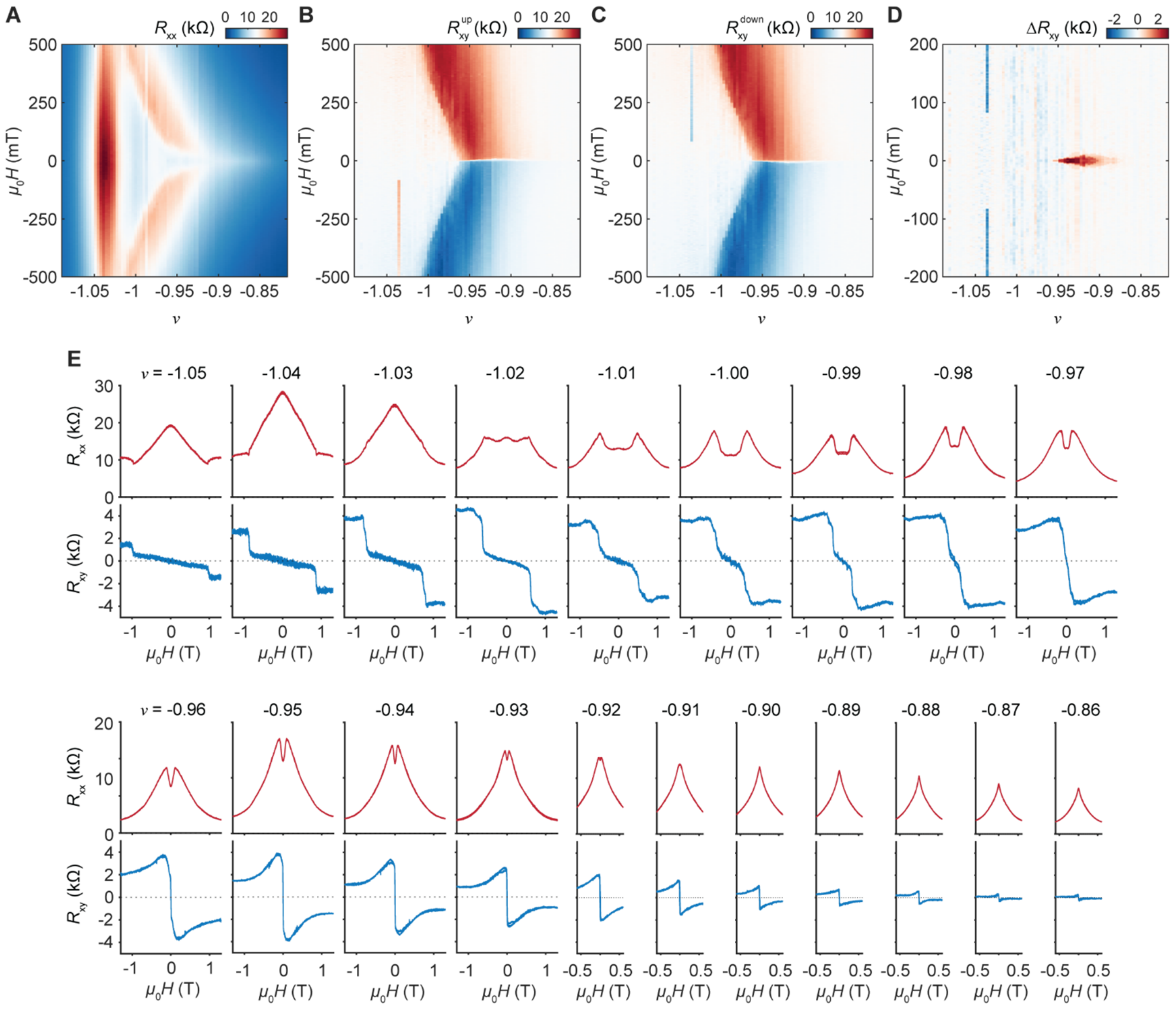

**Fig. S14. Filling dependence of anomalous Hall effect near $\nu$ = -1.** (**A**) $R_{xx}$ as a function of filling factor $\nu$ and magnetic field $\mu_0H$ in the vicinity of $\nu$ = -1. (**B**, **C**) $R_{xy}$ measured during upward (**B**) and downward (**C**) magnetic-field sweeps over the same range. (**D**) Hysteretic component of the Hall response, $\Delta R_{xy}$, obtained from the difference between the up- and down-sweep data, highlighting a narrow region of magnetic hysteresis centered near $\nu$ = -0.94 at low field. (**E**) Field-sweep line cuts of $R_{xx}$ (top panels, red) and $R_{xy}$ (bottom panels, blue) at selected filling factors from $\nu$ = -1.05 to -0.86.

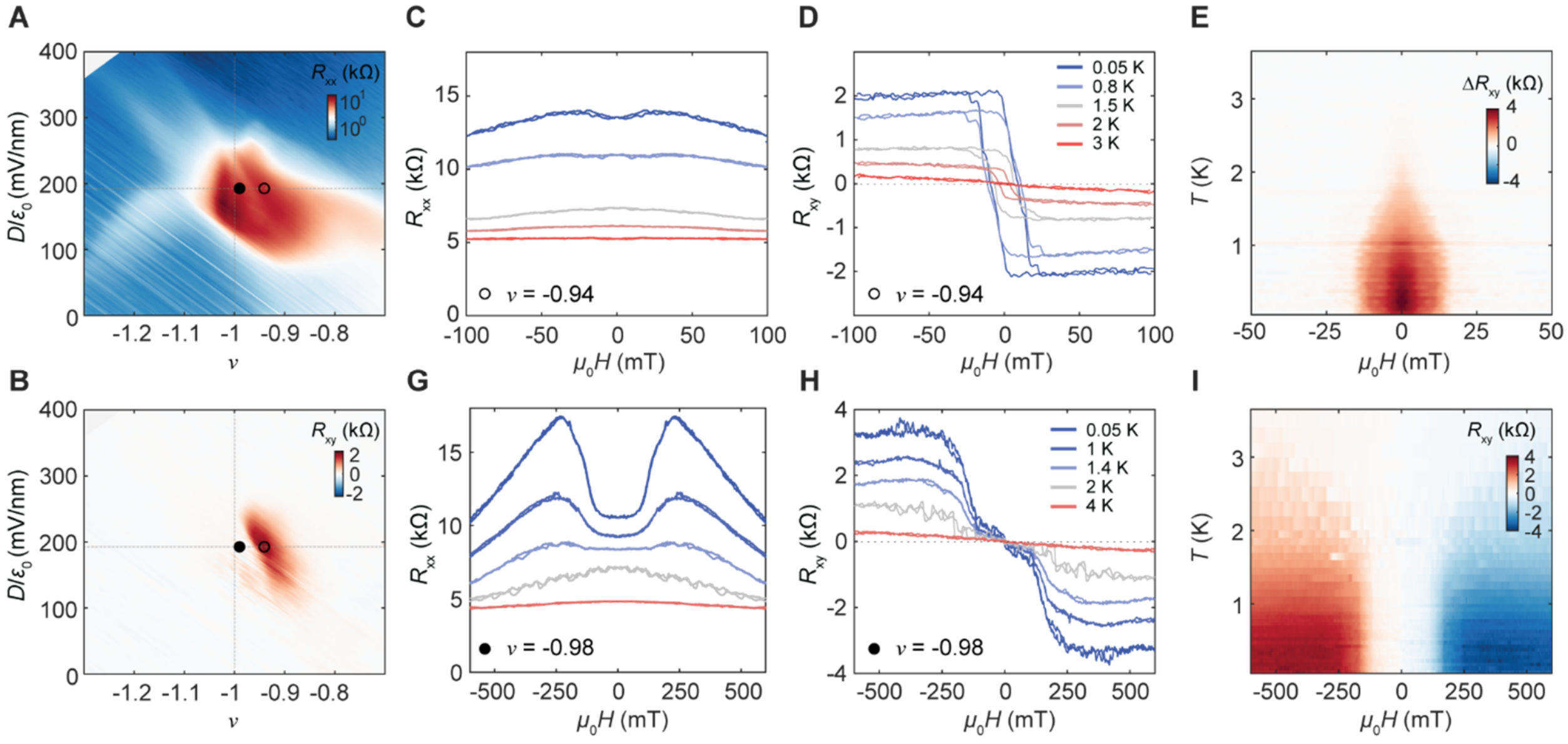

**Fig. S15. Temperature dependence of anomalous Hall effect near $\nu$ = -1.** (**A**, **B**) Maps of longitudinal resistance $R_{xx}$ (**A**) and Hall resistance $R_{xy}$ (**B**) as a function of filling factor $\nu$ and electric field $D/\varepsilon_0$, reproduced from Fig. 2A, B of the main text. The marked points indicate the conditions used for the temperature-dependent measurements. (**C**, **D**) $R_{xx}$ (**C**) and $R_{xy}$ (**D**) measured at $\nu = -0.94$ and $D/\varepsilon_0 = 194$ mV/nm (open circle in **A, B**) while cycling the magnetic field at selected temperatures. The hysteretic anomalous Hall response weakens with increasing temperature. (**E**) Hysteretic component of the Hall resistance ($\Delta R_{xy}$) as a function of magnetic field and temperature, extracted from the difference between up and down magnetic-field sweeps. (**G**, **H**) $R_{xx}$ (**G**) and $R_{xy}$ (**H**) measured at $\nu = -0.98$ and $D/\varepsilon_0 = 194$ mV/nm (filled circle in **A**, **B**) while cycling the magnetic field at selected temperatures. (**I**) $R_{xy}$ as a function of magnetic field and temperature at $\nu = -0.98$.

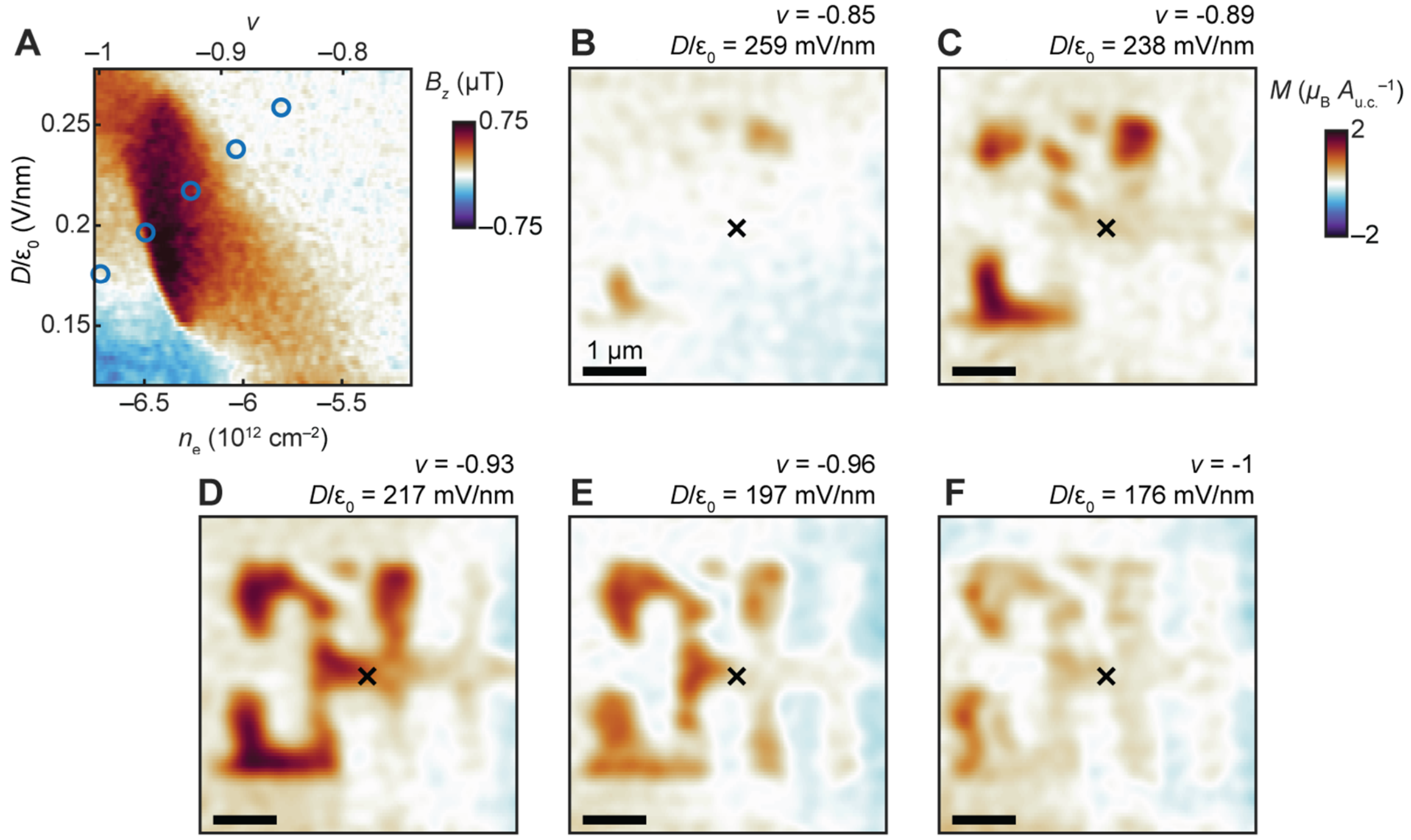

**Fig. S16. Nano SQUID-on-tip measurements.** (**A**) Out-of-plane magnetic field $B_z$ measured at a single point as a function of filling factor $v$ (bottom axis shows the corresponding carrier density $n$) and electric displacement field $D/\varepsilon_0$, measured at 1.6 K at $\mu_0 H$ = 40 mT. (**B**-**F**) Spatial map of out-of-plane magnetization $M$ at selected filling factors and electric fields. Scale bar denotes 1 µm. Cross denotes position where (**A**) was measured.

**References.**